\documentclass[review]{elsarticle}

\usepackage{graphicx}% Include figure files
\usepackage{dcolumn}% Align table columns on decimal point
\usepackage{bm}% bold math
\usepackage{epsfig}
\usepackage{booktabs}
\usepackage{subfigure}
\usepackage{graphics}
\usepackage{amssymb}
\usepackage{amsmath}
\usepackage{array}
\usepackage{color}
\graphicspath{{./pic/}}
\begin{document}

\begin{frontmatter}

\title{Discrete unified gas kinetic scheme for nonlinear convection-diffusion equations}

\author[mymainaddress]{Jinlong Shang}
\author[mymainaddress,mysecondaryaddress]{Zhenhua Chai}
\author[mymainaddress,mythirdaddress]{Huili Wang}
\author[mymainaddress,mysecondaryaddress]{Baochang Shi\corref{mycorrespondingauthor}}
\cortext[mycorrespondingauthor]{Corresponding author}
\ead{shibc@hust.edu.cn}
\address[mymainaddress]{School of Mathematics and Statistics, Huazhong University of Science and Technology, Wuhan 430074, China}
\address[mysecondaryaddress]{Hubei Key Laboratory of Engineering Modeling and Scientific Computing, Huazhong University of Science and Technology, Wuhan 430074, China}
\address[mythirdaddress]{School of Mathematics and Computer Science, Wuhan Textile University, Wuhan, 430073, China}

\begin{abstract}
  In this paper, we develop a discrete unified gas kinetic scheme (DUGKS) for general nonlinear convection-diffusion equation (NCDE), and show that the NCDE can be recovered correctly from the present model through the Chapman-Enskog analysis. We then test the present DUGKS through some classic convection-diffusion equations, and find that the numerical results are in good agreement with analytical solutions and the DUGKS model has a second-order convergence rate. Finally, as a finite-volume method, DUGKS can also adopt the non-uniform mesh. Besides, we performed some comparisons among the DUGKS, finite-volume lattice Boltzmann model (FV-LBM), single-relaxation-time lattice Boltzmann model (SLBM) and multiple-relaxation-time lattice Boltzmann model (MRT-LBM). The results show that the DUGKS model is more accurate than FV-LBM, more stable than SLBM, and almost has the same accuracy as the MRT-LBM. Besides, the using of non-uniform mesh may make DUGKS model more flexible.
\end{abstract}

\begin{keyword}
  Discrete unified gas kinetic scheme \sep Nonlinear convection-diffusion equation \sep Non-uniform mesh
\end{keyword}

\end{frontmatter}

%\linenumbers

\section{\label{sec:level1}Introduction}
  The convection-diffusion equation (CDE) is usually used to describe the physical phenomena where particles, energy or other physical quantities are transferred inside a physical system, and in particular, plays an important role in the field of heat and mass transfer \cite{cussler2009diffusion}. However, as a kind of partial differential equation (PDE), CDE is usually so complicated that it is diffucult to get the analytical solution most of time. With the development of computing power, some numerical methods have been developed to solve CDEs, such as finite-element method \cite{johnson2012numerical}, finite-difference method \cite{thomas2013numerical} and finite-volume method \cite{leveque2002finite}.

  In the past decades, the lattice Boltzmann method (LBM), as a mesoscopic numerical approach, has achieved great success in the simulation of hydrodynamic problems \cite{chen1998lattice,guo2013lattice,succi2015lattice,chen2014critical,dou2013numerical,chai2018comparative,yuan2019dynamic}. On the other hand, the LBM has also been extended to solve the CDEs. Dawson et al. \cite{ponce1993lattice} first proposed a LB model for CDE, but the model cannot give correct CDE. Shi and Guo \cite{shi2009lattice} developed a lattice Bhatnagar-Gross-Krook (LBGK) model to solve the general nonlinear convection-diffusion equations (NCDEs), where an auxiliary moment $\mathbf{C}$ is used to correctly recover the NCDE. However, in their work, the convection term $\mathbf{B}$ should be a function of $\phi$. Chopard \cite{chopard2009lattice} developed a new LB model where a source term related to temporal derivative or spatial derivative is adopted to give correct CDE. We noted that LB models are limited to the isotropic CDEs. To solve the nonlinear anisotropic convection-diffusion equations (NACDEs). The two-relaxation-time (TRT) and multiple-relaxation-time (MRT) LB models are considered by Ginzburg \cite{ginzburg2005equilibrium,ginzburg2005generic,ginzburg2007lattice,ginzburg2012truncation,ginzburg2013multiple}, while in these models, some assumptions on the convection and diffusion terms, and the assumptions may not be satisfied for some special NACDEs. Yoshida and Nagaoka \cite{yoshida2010multiple} also developed a MRT LB model, and did some analysis on different boundary conditions, however, the assumptions were also adopted to recover the CDE. Recently, Chai et al. \cite{chai2016multiple} presented a MRT LB model for general NACDEs without any assumptions on the convection and diffusion terms. Although everything looks perfect, some restrictions still exist in all the above LB models. The first is that the temporal and spatial steps are coupled, causing the selection of parameters to be very limited. The second is that all above LB models must be implemented on uniform grid.

  Recently, Guo et al. \cite{guo2013discrete} proposed the discrete unified gas kinetic scheme (DUGKS) for all Knudsen number flows. The DUGKS combines the advantages of LBM and unified gas kinetic scheme (UGKS). Firstly, as a finite volume scheme, DUGKS can adopt the flexible mesh. Secondly, the DUGKS is more accurate than finite-volume LBM, this is because the evaluation of the flux at cell interface is simplified by employing a transformation of distribution function with collision effect, which has also been used in LBM. Finally, the asymptotic preserving (AP) property still exists in the DUGKS. It should be noted that at the beginning, the DUGKS in Ref. \cite{guo2013discrete} is developed based on the Bhatnagar-Gross-Krook (BGK) collision model \cite{bhatnagar1954model}, and the source term is mot included. Then, Wu et al. \cite{wu2016discrete} developed a DUGKS with a force term for incompressible fluid flows and also presented the non-equilibrium extrapolation (NEE) scheme for DUGKS. Recently, Zhang et al. \cite{zhangchunhua2018discrete} and Yang et al. \cite{yang2019phase} developed the phase-field based DUGKS for two-phase flows, the difference between their two works is that the Chan-Hilliard (CH) equation \cite{cahn1958free,cahn1959free} is considered in Ref. \cite{zhangchunhua2018discrete} while the Allen-Cahn (AC) equation \cite{geier2015conservative} is adopted in Ref. \cite{yang2019phase}. In these works, the DUGKS was used to solve the phase field equations. Huo and Rao \cite{huo2018discrete} uesd the DUGKS to study the solid-liquid phase change problem, in which the energy equation was solved by DUGKS. From above discussion, the DUGKS has been widely used to study single and two-phase flows, and also the phase-field and energy equations. However, it is unclear whether the phase-field and energy equations as some special types of CDEs can be recovered from the DUGKS. Through the DUGKS, we found that the above restrictions of LBM solving CDEs are avoided, perfectly. So whether we can use the present DUGKS model to solve the more general partial differential equations (PDEs)? In this work, we will develop a DUGKS for general NCDEs, and also perform a detailed Chapman-Enskog analysis.

  The rest of the paper is organized as follow. In Sec. \ref{sec:level2}, the DUGKS for the general NCDE is proposed. In Sec. \ref{sec:level3}, through the Chapman-Enskog analysis, the NCDE is recovered correctly from the present DUGKS. In addition, some special cases and distinct characteristics are also discussed. In Sec. \ref{sec:level4}, the accuracy and convergence rate of the DUGKS model are tested through some classic CDEs, and some comparisons among the present DUGKS, finite-volume LB model, LBGK model and MRT LB model are conducted. Finally, some conclusions are given in Sec. \ref{sec:level5}.

\section{\label{sec:level2}The DUGKS model for general NCDEs}
  In this section, we will present a DUGKS for \textit{n}-dimensional NCDE with variable coefficients
  \begin{equation}
    \partial_t \phi+\nabla \cdot \mathbf{B}=\nabla \cdot (\alpha \nabla \cdot \mathbf{D}) + F,
    \label{eq:2_1}
  \end{equation}
  where $\phi$ is a scalar function of position $\mathbf{x}$ and time \textit{t}, $\nabla$ is the gradient operator with respect to the position $\mathbf{x}$ in \textit{n} dimensions. $\mathbf{B}$ and $\mathbf{D}$ are the known convection and diffusion terms, and usually they are related to position $\mathbf{x}$, $\phi$, time \textit{t}. $\alpha$ and $F$ are the diffusion coefficient and source term, respectively.

  Following the idea in the previous work \cite{guo2013discrete}, the DUGKS with D\textit{n}Q\textit{q} lattice (\textit{q} is the number of discrete directions) for the NCDE is considered here. First, the discrete velocity Boltzmann equation (DBE) can be written as
  \begin{equation}
    \frac{\partial f_i}{\partial t} + \mathbf{c}_i \cdot \nabla f_i = \Omega_i + R_i +F_i,
    \label{eq:2_2}
  \end{equation}
  where $f_i = f_i(\mathbf{x},\mathbf{c}_i,t)$ is the particle distribution function with discrete velocity $\mathbf{c}_i$ at time \textit{t} and position $\mathbf{x}$. $\Omega_i = -(f_i - f_i^{eq})/\lambda$ is the Bhatnagar-Gross-Krook (BGK) collision model \cite{shi2009lattice}, $\lambda$ is the relaxation time. $f_i^{eq}$ is the equilibrium distribution function, $R_i$ and $F_i$ are the distribution functions of source term. To derive correctly NCDE (\ref{eq:2_1}) from present DUGKS, the distribution functions $f_i^{eq}$, $R_i$ and $F_i$ are given by
  \begin{equation}
    \begin{split}
      &f_i^{eq}=\omega_i\left[\phi+\frac{\mathbf{c}_i\cdot\mathbf{B}}{c_s^2}+\frac{(c_s^2\mathbf{D}+\mathbf{C}-c_s^2\phi \mathbf{I}):(\mathbf{c}_i \mathbf{c}_i-c_s^2 \mathbf{I})}{2c_s^4}\right],\\
      &R_i=\omega_i\frac{\mathbf{c}_i\cdot(\partial_t\mathbf{B}+\nabla\cdot\mathbf{C})}{c_s^2},\\
      &F_i=\omega_i F,\\
    \end{split}
    \label{eq:2_3}
  \end{equation}
  where $\mathbf{I}$ is the unit matrix, $\mathbf{C}$ is a tensor function which can be set to be $0$ or $\int \mathbf{B}'(\phi)\mathbf{B}'(\phi) d\phi$ \cite{shi2009lattice}. $c_s$ is the so called sound speed related to discrete velocity. $\omega_i$ and $\mathbf{c}_i$ are weight coefficient and discrete velocity, and in different discrete velocity models, they can be defined as\\
  D1Q3:
  \begin{equation}
    \begin{split}
      &\mathbf{c}_i = (0,1,-1)c\\
      &\omega_0=\frac{2}{3}, \omega_1 = \omega_2 = \frac{1}{6}, c_s = \frac{c}{\sqrt{3}},
    \end{split}
    \label{eq:2_4}
  \end{equation}\\
  D2Q9:
  \begin{equation}
    \begin{split}
      &\mathbf{c}_i = \left(\begin{matrix}
        0 & 1 & 0 & -1 & 0 & 1 & -1 & -1 & 1\\
        0 & 0 & 1 & 0 & -1 & 1 & 1 & -1 & -1
      \end{matrix}\right)c\\
      &\omega_0 = \frac{4}{9}, \omega_{1-4} = \frac{1}{9}, \omega_{1-4} = \frac{1}{36}, c_s = \frac{c}{\sqrt{3}},
    \end{split}
    \label{eq:2_5}
  \end{equation}\\
  D3Q15:
  \begin{equation}
    \centering
    \begin{split}
      &\mathbf{c}_i = \left(\begin{matrix}
        0 & 1 & -1 & 0 & 0 & 0 & 0 & 1 & 1 & 1 & -1 & -1 & -1 & -1 & 1\\
        0 & 0 & 0 & 1 & -1 & 0 & 0 & 1 & 1 & -1 & 1 & -1 & -1 & 1 & -1\\
        0 & 0 & 0 & 0 & 0 & 1 & -1 & 1 & -1 & 1 & 1 & -1 & 1 & -1 & -1 
      \end{matrix}\right)c\\
      &\omega_0 = \frac{2}{9}, \omega_{1-6} = \frac{1}{9}, \omega_{7-14} =\frac{1}{72}, c_s = \frac{c}{\sqrt{3}}.
    \end{split}
    \label{eq:2_28}
  \end{equation}\\
  D3Q19:
  \begin{equation}
    \centering
    \begin{split}
      &\mathbf{c}_i = \left(\begin{matrix}
        0 & 1 & -1 & 0 & 0 & 0 & 0 & 1 & -1 & 1 & -1 & 1 & -1 & -1 & 1 & 0 & 0 & 0 & 0\\
        0 & 0 & 0 & 1 & -1 & 0 & 0 & 1 & -1 & -1 & 1 & 0 & 0 & 0 & 0 & 1 & -1 & 1 & -1\\
        0 & 0 & 0 & 0 & 0 & 1 & -1 & 0 & 0 & 0 & 0 & 1 & -1 & 1 & -1 & 1 & -1 & -1 & 1
      \end{matrix}\right)c\\
      &\omega_0 = \frac{1}{3}, \omega_{1-6} = \frac{1}{18}, \omega_{7-18} =\frac{1}{36}, c_s = \frac{c}{\sqrt{3}}.
    \end{split}
    \label{eq:2_6}
  \end{equation}
  Based on the conservation law and Eq. (\ref{eq:2_3}), we have
  \begin{equation}
    \begin{split}
      &\sum_i f_i = \sum_i f_i^{eq} = \phi, \quad \sum_i R_i =0, \quad \sum_i F_i = F,\\
      &\sum_i \mathbf{c}_i f_i^{eq} = \mathbf{B}, \quad \sum_i \mathbf{c}_i R_i = \partial_t \mathbf{B} +\nabla \cdot \mathbf{C}, \quad \sum_i \mathbf{c}_i F_i = 0,\\
      &\sum_i \mathbf{c}_i \mathbf{c}_i f_i^{eq} = c_s^2\mathbf{D} + \mathbf{C}.
    \end{split}
    \label{eq:2_7}
  \end{equation}

  In the DUGKS, we divide the computational domain into a set of control volumes (cells), and $\mathbf{x}_j$ is used to denote the cell $j$. Then, integrating Eq. (\ref{eq:2_2}) over the volume $V_j$ from $t_n$ to $t_{n+1}$, and using the midpoint rule, trapezoidal rule and Taylor expansion for the integration of the flux term at cell interface, collision term and source terms inside the cell, we can obtain
  \begin{equation}
    f_i^{n+1} - f_i^n + \frac{\Delta t}{|V_j|}J^{n+1/2} = \frac{\Delta t}{2}(\Omega_i^{n+1} +\Omega_i^n) + \Delta t(R_i^n + \frac{\Delta t}{2}\partial_t R_i^n) + \Delta t(F_i^n + \frac{\Delta t}{2}\partial_t F_i^n),
    \label{eq:2_8}
  \end{equation}
  where
  \begin{equation}
    J^{n+1/2} = \int_{\partial V_j} (\mathbf{c}_i \cdot \mathbf{n})f_i(\mathbf{x}, \mathbf{c}_i, t_{n+1/2}) d\mathbf{S}
    \label{eq:2_9}
  \end{equation}
  is the flux of cell $j$, $\partial V_j$ and $|V_j|$ are the surface area and volume of cell $j$, $\mathbf{n}$ is the outward unit vector to the surface, $\Delta t = \alpha \frac{\Delta x}{c}$ is the time step and it is only determined by the Courant-Friedrichs-Lewy (CFL) condition ($\alpha$ is the CFL number and lies between 0 and 1). It should be noted that $f_i^n$, $R_i^n$, $F_i^n$ and $\Omega_i^n$ in Eq. (\ref{eq:2_8}) are the cell-averaged values of the distribution functions and collision term, respectively, i.e.,
  \begin{equation}
    A_i^n = \frac{1}{|V_j|} \int_{V_j} A_i(\mathbf{x}_j, \mathbf{c}_i, t_n) d\mathbf{x}, \qquad A\in\left\{f,R,F,\Omega\right\}
    \label{eq:2_10}
  \end{equation}

  Because the collision term $\Omega_i^{n+1}$ involves the unknown variables at $t_{n+1}$, thus the evolution equation Eq. (\ref{eq:2_8}) is implicit scheme. In order to remove the implicity, a new distribution function is adopted,
  \begin{equation}
    \tilde{f}_i = f_i -\frac{\Delta t}{2} \Omega_i = \frac{2\lambda + \Delta t}{2\lambda}f_i - \frac{\Delta t}{2\lambda}f_i^{eq}.
    \label{eq:2_11}
  \end{equation}
  Then Eq. (\ref{eq:2_8}) can be rewritten as
  \begin{equation}
    \tilde{f}_i^{n+1} = \tilde{f}_i^{+,n} -\frac{\Delta t}{|V_j|}J^{n+1/2} + \Delta t(R_i^n + \frac{\Delta t}{2}\partial_t R_i^n) + \Delta t(F_i^n + \frac{\Delta t}{2}\partial_t F_i^n),
    \label{eq:2_12}
  \end{equation}
  where
  \begin{equation}
    \tilde{f}_i^+ = \frac{2\lambda - \Delta t}{2\lambda + \Delta t} \tilde{f}_i + \frac{2\Delta t}{2\lambda + \Delta t}f_i^{eq}.
    \label{eq:2_13}
  \end{equation}
  Based on Eqs. (\ref{eq:2_7}) and (\ref{eq:2_11}), the conserved variable $\phi$ can be computed by $\phi = \sum_i \tilde{f}_i$. With this fact, we only need to track the distribution function $\tilde{f}_i$ instead of $f_i$ in practical computation. Besides, from the computational point of view, if we use $(R_i^n-R_i^{n-1})/\Delta t$ and $(F_i^n-F_i^{n-1})/\Delta t$ to evaluate the values of $\partial_t R_i^n$ and $\partial_t F_i^n$, Eq. (\ref{eq:2_12}) would become an explicit format. However, if $(R_i^{n+1}-R_i^n)/\Delta t$ and $(F_i^{n+1}-F_i^n)/\Delta t$ are used to estimate $\partial_t R_i^n$ and $\partial_t F_i^n$, we can rewrite Eq. (\ref{eq:2_12}) as 
  \begin{equation}
    \tilde{f}_i^{n+1} = \tilde{f}_i^{+,n} -\frac{\Delta t}{|V_j|}J^{n+1/2},
    \label{eq:2_24}
  \end{equation}
  with
  \begin{equation}
    \begin{split}
      &\tilde{f}_i = f_i -\frac{\Delta t}{2} \Omega_i - \frac{\Delta t}{2} R_i - \frac{\Delta t}{2} F_i,\\
      &\tilde{f}_i^+ = \frac{2\lambda - \Delta t}{2\lambda + \Delta t} \tilde{f}_i + \frac{2\Delta t}{2\lambda + \Delta t}f_i^{eq} + \frac{2 \lambda \Delta t}{2\lambda + \Delta t}R_i + \frac{2 \lambda \Delta t}{2\lambda + \Delta t}F_i.
    \end{split}
    \label{eq:2_25}
  \end{equation}
  This method is the same as Ref. \cite{wu2016discrete}. If necessary, we can also use $(F_i^n-F_i^{n-1})/\Delta t$ to estimate $\partial_t F_i^n$ and use $(R_i^{n+1}-R_i^n)/\Delta t$ to estimate $\partial_t R_i^n$.

\begin{figure}
    \centering
    \includegraphics[scale=0.7]{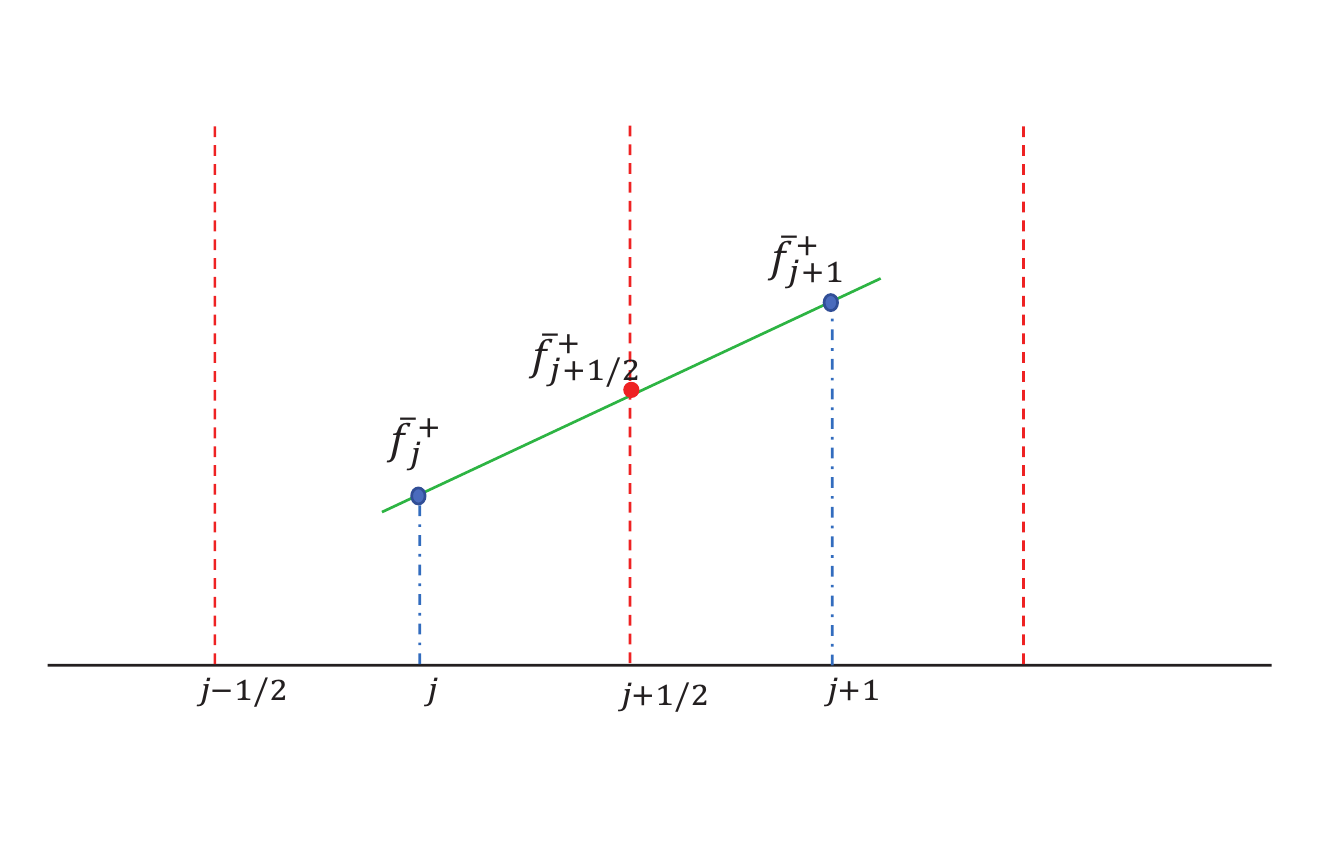}
    \caption{Illustration of one-dimensional cell geometry}
    \label{Fig2_1}
  \end{figure}

  Now, the key ingredient in updating $\tilde{f}_i$ according to Eq. (\ref{eq:2_12}) is to evaluate the flux $J^{n+1/2}$. From Eq. (\ref{eq:2_9}), we can see that the flux $J^{n+1/2}$ is only determined by the original distribution $f_i$ at time $t+\frac{\Delta t}{2}$. In order to compute $f_i(\mathbf{x},\mathbf{c}_i,t_{n+1/2})$, we integrate the discrete velocity Boltzmann equation within a half time step $h=\Delta t/2$ along the characteristic line with the point ($\mathbf{x}_b$) located at the cell interface ($\mathbf{x}_b = \mathbf{x}_{j+1/2}$ in the one-dimensional case, see Fig. \ref{Fig2_1}),
  \begin{equation}
    \begin{split}
      &f_i(\mathbf{x}_b,\mathbf{c}_i,t_n+h) - f_i(\mathbf{x}_b-\mathbf{c}_ih,\mathbf{c}_i,t_n) = \frac{h}{2}[\Omega_i(\mathbf{x}_b,\mathbf{c}_i,t_n+h) \\
      &+ \Omega_i(\mathbf{x}_b-\mathbf{c}_ih,\mathbf{c}_i,t_n)] + h[R_i(\mathbf{x}_b-\mathbf{c}_ih,\mathbf{c}_i,t_n) + \frac{h}{2}D_iR_i(\mathbf{x}_b-\mathbf{c}_ih,\mathbf{c}_i,t_n)] \\
      &+ h[F_i(\mathbf{x}_b-\mathbf{c}_ih,\mathbf{c}_i,t_n) + \frac{h}{2}D_iF_i(\mathbf{x}_b-\mathbf{c}_ih,\mathbf{c}_i,t_n)],
    \end{split}
    \label{eq:2_14}
  \end{equation}
  where the trapezoidal rule and Taylor expansion are applied to evaluate the collision term and source terms, $D_i = \partial_t + \mathbf{c}_i \cdot \nabla$. Then, similar to the treatment in Eq. (\ref{eq:2_11}), another distribution function $\bar{f}$ is introduced to remove the implicity of Eq. (\ref{eq:2_14}).
  \begin{equation}
    \bar{f}_i = f_i - \frac{h}{2}\Omega_i = \frac{2\lambda + h}{2\lambda} - \frac{h}{2\lambda} f_i^{eq}.
    \label{eq:2_15}
  \end{equation}
  As a result, the implicit formulation Eq. (\ref{eq:2_14}) can be rewritten by
  \begin{equation}
    \begin{split}
      &\bar{f}_i(\mathbf{x}_b,\mathbf{c}_i,t_n+h) = \bar{f}_i^+(\mathbf{x}_b-\mathbf{c}_ih,\mathbf{c}_i,t_n) + h[R_i(\mathbf{x}_b-\mathbf{c}_ih,\mathbf{c}_i,t_n) \\
      &+ \frac{h}{2}D_iR_i(\mathbf{x}_b-\mathbf{c}_ih,\mathbf{c}_i,t_n)] + h[F_i(\mathbf{x}_b-\mathbf{c}_ih,\mathbf{c}_i,t_n) + \frac{h}{2}D_iF_i(\mathbf{x}_b-\mathbf{c}_ih,\mathbf{c}_i,t_n)],
    \end{split}
    \label{eq:2_16}
  \end{equation}
  where
  \begin{equation}
    \bar{f}_i^+ = \frac{2\lambda - h}{2\lambda +h}\bar{f}_i + \frac{2h}{2\lambda + h}f_i^{eq}.
    \label{eq:2_17}
  \end{equation}
  Here, we can adopt $[R_i(\mathbf{x}_b,\mathbf{c}_i,t_n+h)-R_i(\mathbf{x}_b-\mathbf{c}_ih,\mathbf{c}_i,t_n)]/h$ and $[F_i(\mathbf{x}_b,\mathbf{c}_i,t_n+h)-F_i(\mathbf{x}_b-\mathbf{c}_ih,\mathbf{c}_i,t_n)]/h$ to evaluate the values of $D_iR_i(\mathbf{x}_b-\mathbf{c}_ih,\mathbf{c}_i,t_n)$ and $D_iF_i(\mathbf{x}_b-\mathbf{c}_ih,\mathbf{c}_i,t_n)$, and Eq. (\ref{eq:2_16}) can be rewritten as
  \begin{equation}
    \bar{f}_i(\mathbf{x}_b,\mathbf{c}_i,t_n+h) = \bar{f}_i^+(\mathbf{x}_b-\mathbf{c}_ih,\mathbf{c}_i,t_n),
    \label{eq:2_26}
  \end{equation}
  with
  \begin{equation}
    \begin{split}
      &\bar{f}_i = f_i - \frac{h}{2}\Omega_i - \frac{h}{2}R_i -\frac{h}{2}F_i,\\
      &\bar{f}_i^+ = \frac{2\lambda - h}{2\lambda + h} \bar{f}_i + \frac{2h}{2\lambda + h}f_i^{eq} + \frac{2 \lambda h}{2\lambda + h}R_i + \frac{2 \lambda h}{2\lambda + h}F_i.
    \end{split}
    \label{eq:2_27}
  \end{equation}
  The method is the same as Ref. \cite{wu2016discrete}. If the source term is a nonlinear function of the variable $\phi$, we can use the explicit difference method to avoid solving the nonlinear equations, but, the finite-difference scheme for gradient term would destroy the locality of the DUGKS.

  Actually, based on the previous works \cite{shi2009lattice,chai2016multiple}, we can rewrite the evolution equation Eq. (\ref{eq:2_16}) as
  \begin{equation}
    \begin{split}
      &\bar{f}_i(\mathbf{x}_b,\mathbf{c}_i,t_n+h) = \bar{f}_i^+(\mathbf{x}_b-\mathbf{c}_ih,\mathbf{c}_i,t_n) + \frac{2\lambda h}{2\lambda + h}[R_i(\mathbf{x}_b-\mathbf{c}_ih,\mathbf{c}_i,t_n) \\
      &+ \frac{h}{2}\partial_t R_i(\mathbf{x}_b-\mathbf{c}_ih,\mathbf{c}_i,t_n)] + h[F_i(\mathbf{x}_b-\mathbf{c}_ih,\mathbf{c}_i,t_n) + \frac{h}{2}\partial_t F_i(\mathbf{x}_b-\mathbf{c}_ih,\mathbf{c}_i,t_n)].
      \label{eq:2_18}
    \end{split}
  \end{equation}
  This evolution equation can avoid calculating the gradient term, and we will explain why we can do this in next section. Now, we focus on the computation of the distribution function $\bar{f}_i^+(\mathbf{x}_b-\mathbf{c}_ih,\mathbf{c}_i,t_n)$, $R_i(\mathbf{x}_b-\mathbf{c}_ih,\mathbf{c}_i,t_n)$ and $F_i(\mathbf{x}_b-\mathbf{c}_ih,\mathbf{c}_i,t_n)$.
  
  With the Taylor expansion, the cell interface $\mathbf{x}_b$, the distribution function $\bar{f}_i^+(\mathbf{x}_b-\mathbf{c}_ih,\mathbf{c}_i,t_n)$, $R_i(\mathbf{x}_b-\mathbf{c}_ih,\mathbf{c}_i,t_n)$ and $F_i(\mathbf{x}_b-\mathbf{c}_ih,\mathbf{c}_i,t_n)$ can be approximated as
  \begin{equation}
    \begin{split}
      &\bar{f}_i^+(\mathbf{x}_b-\mathbf{c}_ih,\mathbf{c}_i,t_n) = \bar{f}_i^+(\mathbf{x}_b,\mathbf{c}_i,t_n) - h\mathbf{c}_i \cdot \nabla \bar{f}_i^+(\mathbf{x}_b,\mathbf{c}_i,t_n),\\
      &R_i(\mathbf{x}_b-\mathbf{c}_ih,\mathbf{c}_i,t_n) = R_i(\mathbf{x}_b,\mathbf{c}_i,t_n) - h\mathbf{c}_i \cdot \nabla R_i(\mathbf{x}_b,\mathbf{c}_i,t_n),\\
      &F_i(\mathbf{x}_b-\mathbf{c}_ih,\mathbf{c}_i,t_n) = F_i(\mathbf{x}_b,\mathbf{c}_i,t_n) - h\mathbf{c}_i \cdot \nabla F_i(\mathbf{x}_b,\mathbf{c}_i,t_n),
    \end{split}
    \label{eq:2_19}
  \end{equation}
  where the distribution functions at $\mathbf{x}_b$ and the gradient terms can be approximated by linear interpolations, respectively. For example, as shown in Fig. \ref{Fig2_1}, in one-dimensional case, the reconstructions become
    \begin{equation}
      \begin{split}
        &\nabla \bar{f}_i^+(x_{j+1/2},\mathbf{c}_i,t_n) = \frac{\bar{f}_i^+(x_{j+1},\mathbf{c}_i,t_n) - \bar{f}_i^+(x_j,\mathbf{c}_i,t_n)}{x_{j+1} - x_j},\\
        &\bar{f}_i^+(x_{j+1/2},\mathbf{c}_i,t_n) = \bar{f}_i^+(x_j,\mathbf{c}_i,t_n) + (x_{j+1/2} - x_j) \nabla \bar{f}_i^+(x_{j+1/2},\mathbf{c}_i,t_n).
      \end{split}
    \label{eq:2_20}
  \end{equation}
  The estimations of $R_i(\mathbf{x}_b,\mathbf{c}_i,t_n)$, $F_i(\mathbf{x}_b,\mathbf{c}_i,t_n)$, $\nabla R_i(\mathbf{x}_b,\mathbf{c}_i,t_n)$ and $\nabla F_i(\mathbf{x}_b,\mathbf{c}_i,t_n)$ are similar to Eq. (\ref{eq:2_20}). From Eqs. (\ref{eq:2_15}) and (\ref{eq:2_18}), we can obtain the conserved variable at the cell interface $\mathbf{x}_b$, $\phi(\mathbf{x}_b,t_n + h) = \sum_i \bar{f}_i(\mathbf{x}_b,\mathbf{c}_i,t_n + h)$.

  Besides, we can get the relationship of distribution functions $f_i$, $f_i^{eq}$, $\bar{f}_i$, $\tilde{f}_i$, $\bar{f}_i^+$ and $\tilde{f}_i^+$ from Eqs. (\ref{eq:2_11}), (\ref{eq:2_13}), (\ref{eq:2_15}) and (\ref{eq:2_17}),
  \begin{equation}
    \bar{f}_i^+ = \frac{2\lambda-h}{2\lambda + \Delta t} \tilde{f}_i + \frac{3h}{2\lambda + \Delta t} f_i^{eq},
    \label{eq:2_21}
  \end{equation}
  \begin{equation}
    f_i = \frac{2\lambda}{2\lambda+h} \bar{f_i}+\frac{h}{2\lambda + h} f_i^{eq},
    \label{eq:2_22}
  \end{equation}
  \begin{equation}
    \tilde{f}_i^+ = \frac{4}{3}\bar{f}_i^+ - \frac{1}{3}\tilde{f}_i.
    \label{eq:2_23}
  \end{equation}

  The update of conserved variable $\phi$ in one time step of the present DUGKS can be summarized as follows:
  \begin{equation*}
    \begin{split}
      \phi(\mathbf{x}_j,t),F(\mathbf{x}_j,t) \quad &\xrightarrow{(\ref{eq:2_3})} \quad \tilde{f}_i(\mathbf{x}_j,t) \,,\, R_i(\mathbf{x}_j,t) \,,\, F_i(\mathbf{x}_j,t)\\
      &\xrightarrow{(\ref{eq:2_13}),(\ref{eq:2_21})}\quad \tilde{f}_i^+(\mathbf{x}_j,t) \,,\, \bar{f}_i^+(\mathbf{x}_j,t)\\
      &\xrightarrow{(\ref{eq:2_20})} \quad \bar{f}_i^+(\mathbf{x}_b,t) \,,\,  R_i(\mathbf{x}_b,t) \,,\,  F_i(\mathbf{x}_b,t)\\
      &\xrightarrow{(\ref{eq:2_18}),(\ref{eq:2_19})} \quad \bar{f}_i(\mathbf{x}_b,t+h) \quad \xrightarrow \quad \phi(\mathbf{x}_b,t+h)\\
      &\xrightarrow{(\ref{eq:2_22})} \quad f_i(\mathbf{x}_b,t+h) \quad \xrightarrow{(\ref{eq:2_9})} \quad J(\mathbf{x}_b,t+h)\\
      &\xrightarrow{(\ref{eq:2_12})} \quad \tilde{f}_i(\mathbf{x}_j,t+\Delta t) \quad \xrightarrow \quad \phi(\mathbf{x}_j,t + \Delta t) \,,\, F(\mathbf{x}_j,t+\Delta t)
    \end{split}
  \end{equation*}

\section{\label{sec:level3}The Chapman-Enskog Analysis}
  In this part, the present DUGKS for NCDE is analyzed through the Chapman-Enskog (CE) analysis. In general, the using of Chapman-Enskog analysis in LBM is to recover the macroscopic equations from evolution equations \cite{shi2009lattice,chai2016multiple}. However, we found that the using of CE analysis in DUGKS is to recover the macroscopic equations from DBE \cite{zhangchunhua2018discrete}. The results of the two ways are the same, such as equilibrium distribution function, moment conditions and so on. In the following, the NCDE will be exactly recovered from Eq. (\ref{eq:2_2}) and (\ref{eq:2_18}), respectively.

  Firstly, we expand the distribution functions $f_i$, $R_i$, $F_i$, the derivatives of time and space as
  \begin{equation}
    \begin{split}
      &f_i = f_i^{(0)} + \epsilon f_i^{(1)} +\epsilon^2 f_i^{(2)}, \quad  R_i = \epsilon R_i^{(1)} + \epsilon^2 R_i^{(2)}, \quad F_i = \epsilon F_i^{(1)} + \epsilon^2 F_i^{(2)},\\
      &\partial_t = \epsilon \partial_{t_1} + \epsilon^2 \partial_{t_2}, \quad \nabla = \epsilon \nabla_1,
    \end{split}
    \label{eq:3_1}
  \end{equation}
  where $\epsilon$ is a small parameter and keeps the same order of the Knudsen number.

  Substituting Eq. (\ref{eq:3_1}) into Eq. (\ref{eq:2_2}), some equations at different orders of $\epsilon$ are obtained,
  \begin{equation}
    \begin{split}
      &O(\epsilon^0):\quad f_i^{(0)}=f_i^{eq},\\
      &O(\epsilon^1):\quad \partial_{t_1}f_i^{(0)}+\mathbf{c}_i\cdot\nabla_1 f_i^{(0)}=-\frac{1}{\lambda}f_i^{(1)}+R_i^{(1)}+F_i^{(1)},\\
      &O(\epsilon^2):\quad \partial_{t_1}f_i^{(1)}+\partial_{t_2}f_i^{(0)}+\mathbf{c}_i\cdot\nabla_1 f_i^{(1)}=-\frac{1}{\lambda}f_i^{(2)} +R_i^{(2)} + F_i^{(2)}.
    \end{split}
    \label{eq:3_2}
  \end{equation}
  Summing Eq. (\ref{eq:3_2}) over $i$, we can get
  \begin{equation}
    \begin{split}
      &O(\epsilon^0):\quad \sum_if_i^{(0)}=\sum_if_i^{eq},\\
      &O(\epsilon^1):\quad \partial_{t_1}\sum_if_i^{(0)}+\nabla_1\cdot\sum_i\mathbf{c}_if_i^{(0)}=-\frac{1}{\lambda}\sum_if_i^{(1)}+\sum_iR_i^{(1)}+\sum_iF_i^{(1)},\\
      &O(\epsilon^2):\quad \partial_{t_1}\sum_if_i^{(1)}+\partial_{t_2}\sum_if_i^{(0)}+\nabla_1\cdot\sum_i\mathbf{c}_if_i^{(1)}=-\frac{1}{\lambda}\sum_if_i^{(2)} + \sum_i R_i^{(2)} + \sum_i F_i^{(2)}.
    \end{split}
    \label{eq:3_3}
  \end{equation}
  From Eq. (\ref{eq:3_2}), we can obtain $\sum_i \mathbf{c}_if_i^{(1)}$,
  \begin{equation}
    \sum_i \mathbf{c}_i f_i^{(1)}=-\lambda(\partial_{t_1}\sum_i \mathbf{c}_i f_i^{(0)}+\nabla_1 \cdot \sum_i \mathbf{c}_i \mathbf{c}_i f_i^{(0)} - \sum_i \mathbf{c}_i R_i^{(1)} - \sum_i \mathbf{c}_i F_i^{(1)}).
    \label{eq:3_4}
  \end{equation}
  Substituting Eq. (\ref{eq:3_4}) into Eq. (\ref{eq:3_3}) yields
  \begin{equation}
    \begin{split}
      &O(\epsilon^0):\quad \sum_if_i^{(0)}=\sum_if_i^{eq},\\
      &O(\epsilon^1):\quad \partial_{t_1}\sum_if_i^{(0)}+\nabla_1\cdot\sum_i\mathbf{c}_if_i^{(0)}=-\frac{1}{\lambda}\sum_if_i^{(1)}+\sum_iR_i^{(1)}+\sum_iF_i^{(1)},\\
      &O(\epsilon^2):\quad \partial_{t_1}\sum_if_i^{(1)}+\partial_{t_2}\sum_if_i^{(0)}+\nabla_1\cdot[-\lambda(\partial_{t_1}\sum_i \mathbf{c}_i f_i^{(0)}+\nabla_1 \cdot \sum_i \mathbf{c}_i \mathbf{c}_i f_i^{(0)}\\
      &\qquad \qquad -\sum_i \mathbf{c}_i R_i^{(1)} - \sum_i \mathbf{c}_i F_i^{(1)})]=-\frac{1}{\lambda} \sum_i f_i^{(2)} + \sum_i R_i^{(2)} + \sum_i F_i^{(2)}.
    \end{split}
    \label{eq:3_5}
  \end{equation}
  Using the conditions in Eq. (\ref{eq:2_7}), we have
  \begin{equation}
    \begin{split}
      &O(\epsilon^1):\quad \partial_{t_1}\phi+\nabla_1\cdot\mathbf{B}=F^{(1)},\\
      &O(\epsilon^2):\quad \partial_{t_2}\phi=\nabla_1\cdot[\alpha\nabla_1 \cdot \mathbf{D}] + F^{(2)}.
    \end{split}
    \label{eq:3_6}
  \end{equation}
  Combining above equations at the orders of $O(\epsilon^1)$ and $O(\epsilon^2)$, and taking $\alpha = c_s^2 \lambda$, we can get the Eq. (\ref{eq:2_1}).

  In addition, we redesign Eq. (\ref{eq:2_16}) to Eq. (\ref{eq:2_18}), and show that Eq. (\ref{eq:2_1}) can be exactly recovered from Eq. (\ref{eq:2_18}). We expand the distribution functions $\bar{f}_i$, $R_i$, $F_i$, the derivatives of time and space, the same as Eq. (\ref{eq:3_1}),
  \begin{equation}
    \begin{split}
      &\bar{f}_i = \bar{f}_i^{(0)} + \epsilon \bar{f}_i^{(1)} +\epsilon^2 \bar{f}_i^{(2)}, \quad  R_i = \epsilon R_i^{(1)} + \epsilon^2 R_i^{(2)}, \quad F_i = \epsilon F_i^{(1)} + \epsilon^2 F_i^{(2)},\\
      &\partial_t = \epsilon \partial_{t_1} + \epsilon^2 \partial_{t_2}, \quad \nabla = \epsilon \nabla_1.
    \end{split}
    \label{eq:3_7}
  \end{equation}
  By applying Taylor expansion to Eq. (\ref{eq:2_18}), we have
  \begin{equation}
    D_i \bar{f}_i + \frac{h}{2}D_i^2\bar{f}_i = -\frac{2}{2\lambda + h}(\bar{f}_i - f_i^{eq}) + \frac{2\lambda}{2\lambda + h}(R_i + \frac{h}{2}\partial_t R_i) + (F_i + \frac{h}{2}\partial_t F_i),
    \label{eq:3_8}
  \end{equation}
  where $D_i = \epsilon D_{1i} + \epsilon^2\partial_{t_2}$ and $D_{1i} = \partial_{t_1} + \mathbf{c}_i \cdot \nabla_1$.

  Substituting Eq. (\ref{eq:3_7}) into Eq. (\ref{eq:3_8}), we can derive the following equations at different orders of $\epsilon$,
  \begin{equation}
    \begin{split}
      O(\epsilon^0): \quad &\bar{f}_i^{(0)} = f_i^{eq},\\
      O(\epsilon^1): \quad &D_{1i}\bar{f}_i^{(0)} = -\frac{2}{2\lambda + h}\bar{f}_i^{(1)} + \frac{2\lambda}{2\lambda + h}R_i^{(1)} + F_i^{(1)},\\
      O(\epsilon^2): \quad &\partial_{t_2}\bar{f}_i^{(0)} + D_{1i}\bar{f}_i^{(1)} + \frac{h}{2}D_{1i}^2\bar{f}_i^{(0)} = -\frac{2}{2\lambda + h}\bar{f}_i^{(2)} \\
      &+ \frac{2\lambda}{2\lambda + h}(R_i^{(2)} + \frac{h}{2}\partial_{t_1}R_i^{(1)}) + (F_i^{(2)} + \frac{h}{2}\partial_{t_1}F_i^{(1)}).
    \end{split}
    \label{eq:3_9}
  \end{equation}
  Summing Eq. (\ref{eq:3_9}) over $i$ and using Eq. (\ref{eq:2_7}), we can get
  \begin{equation}
    \begin{split}
      &O(\epsilon^1): \quad \partial_{t_1} \phi + \nabla_1 \cdot \mathbf{B} = F^{(1)},\\
      &O(\epsilon^2): \quad \partial_{t_2} \phi + \frac{2\lambda}{2\lambda + h}\nabla_1 \cdot \sum_i\mathbf{c}_i\bar{f}_i^{(1)} + \frac{\lambda h}{2\lambda + h}\nabla_1 (\partial_{t_1}\mathbf{B} + \nabla_1 \cdot \mathbf{C}) = F^{(2)}.
    \end{split}
    \label{eq:3_10}
  \end{equation}
  From Eq. (\ref{eq:3_2}), we have
  \begin{equation}
    \begin{split}
      \sum_i \mathbf{c}_i \bar{f}_i^{(1)} &= -\frac{2\lambda + h}{2} \sum_i \mathbf{c}_i (D_{1i}\bar{f}_i^{(0)} - \frac{2\lambda}{2\lambda + h}R_i^{(1)} - F_i^{(1)})\\
      &= -\frac{2\lambda + h}{2}[\nabla_1 \cdot c_s^2\mathbf{D} + \frac{h}{2\lambda + h}(\partial_{t_1}\mathbf{B} + \nabla_1 \cdot \mathbf{C})].
    \end{split}
    \label{eq:3_11}
  \end{equation}
  Then, substituting Eq. (\ref{eq:3_11}) into Eq. (\ref{eq:3_10}), we can obtain Eq. (\ref{eq:3_6}). Similarly, combining equations at the orders $O(\epsilon^1)$ and $O(\epsilon^2)$, and taking $\alpha = c_s^2 \lambda$, the NCDE (\ref{eq:2_1}) is correctly recovered from Eq. (\ref{eq:2_18}). 

  \emph{Remark 1.}As a model only include one relaxation time, DUGKS is different from single-relaxation-time lattice Boltzmann model (SLBM). The spatial and temporal step in DUGKS are decoupled, the temporal step in DUGKS is determined by  Courant-Friedrichs-Lewy(CFL) condition($\Delta t = \alpha \frac{\Delta x}{c}$). In SLBM, however, the spatial and temporal steps are coupled through $\Delta \mathbf{x} = \mathbf{c}_i \Delta t$. Therefore, the restriction in SLBM does not exist in DUGKS.

  \emph{Remark 2.}As a finite volume scheme, the present DUGKS is different from finite-volume lattice Boltzmann Method (FV-LBM). In DUGKS, the flux $J$ is appropriated by $f_i(t_{n+1/2})$ instead of $f_i(t_n)$ in FV-LBM. The analysis in Refs. \cite{patil2013chapman,ubertini2004recent} show that the FV-LBM may suffer from severe numerical dissipation.

  \emph{Remark 3.}The tensor function $\mathbf{C}$ in Eq. (\ref{eq:2_3}) is an auxiliary-moment. If $\mathbf{B}$ is the function of $\phi$, $u$, $\mathbf{x}$ and \textit{t}, we can define $\mathbf{C} = 0$ so that $R_i = \omega_i\frac{\mathbf{c}_i \cdot \partial_t \mathbf{B}}{c_s^2}$. If $\mathbf{B}$ is only a function of $\phi$, we can define $\mathbf{C} = \int \mathbf{B}'(\phi)\mathbf{B}'(\phi)d\phi$ so that $R_i = \omega_i\frac{\mathbf{c}_i\cdot \mathbf{B}'(\phi) F}{c_s^2}$. In the second case, we do not have to calculate the temporal derivative, and additionally, the equilibrium distribution function $f_i^{eq}$ can also be simplified by $f_i^{eq} = \omega_i(\phi + \frac{\mathbf{c}_i \cdot \mathbf{B}}{c_s^2})$ if $\mathbf{D}$ is just a function of $\phi$, too. Furthermore, when the linear equilibrium distribution function $f_i^{eq} = \omega_i(\phi + \frac{\mathbf{c}_i \cdot \mathbf{B}}{c_s^2})$ is considered, we can adopt the DdQ2d+1 discrete velocity model, for instance, D1Q3, D2Q5 and D3Q7.

\section{\label{sec:level4}Numerical results and discussion}
  In this part, some examples, including isotropic CDE with a constant velocity, Burgers-Fisher equation, the nonlinear heat conduction equation (NHCE), Gaussian hill problem and CDE with nonlinear convection and diffusion terms, are adopted to test the accuracy and stability of the present DUGKS. In our simulations, the distribution function $\tilde{f}_i$ is initialized by the equilibrium distribution function $f_i^{eq}$, i.e., $\tilde{f}_i(\mathbf{x},t_0) = f_i^{eq}(\mathbf{x},t_0)$. Unless otherwise stated, the non-equilibrium extrapolation scheme \cite{wu2016discrete} is used to treat the boundary conditions. The following global relative error ($GRE$) is used to measure the accuracy of the present DUGKS,
  \begin{equation}
    GRE = \frac{\sum_i |\phi_a(\mathbf{x},t) - \phi_n(\mathbf{x},t)|}
    {\sum_i |\phi_a(\mathbf{x},t)|},
    \label{eq:4_1_1}
  \end{equation}
  where $\phi_a$ and $\phi_n$ are the analytical and numerical solutions. In addition, to obtain stable results with present DUGKS, the CFL condition number should be less than 1.

  \textbf{Example 4.1} Two-dimensional isotropic CDE with a constant velocity can be expressed as
  \begin{equation}
    \partial_t \phi + \partial_x(u_x \phi) + \partial_y(u_y \phi) = \alpha (\partial_{xx} \phi + \partial_{yy} \phi) + F,
    \label{eq:4_1_2}
  \end{equation}
  where $u_x$ and $u_y$ are constants, and set to be 0.1, $\alpha$ is the diffusion coefficient. $F$ is the source term, and is given by
  \begin{equation}
    F = \exp [(1 - 2\pi^2\alpha) t]  \left\{\sin[\pi(x + y)] + \pi(u_x + u_y)
    \cos[\pi(x +y)]\right\}.
    \label{eq:4_1_3}
  \end{equation}
  Under the periodic boundary and following initial conditions,
  \begin{equation}
    \phi(x,y,t=0) = \sin[\pi(x + y)], \quad (x,y) \in [0,2]\times[0,2],
    \label{eq:4_1_4}
  \end{equation}
  the solution of the problem can be expressed as
  \begin{equation}
    \phi(x,y,t) = \exp[(1-2\pi^2\alpha)t]\sin[\pi(x + y)].
    \label{eq:4_1_5}
  \end{equation}
  When the present DUGKS is used to study this problem, the functions $\mathbf{B}$, $\mathbf{C}$ and $\mathbf{D}$ are given by $\mathbf{B} = \phi \mathbf{u}$ with $\mathbf{u} = (u_x , u_y)^T$, $\mathbf{C} = \phi \mathbf{u}\mathbf{u}$ and $\mathbf{D} = \phi \mathbf{I}$.

\begin{figure}[ht]
    \centering
    \subfigure[]
    {
      \label{Fig4_1_1_a}
      \includegraphics[width=0.7\textwidth]{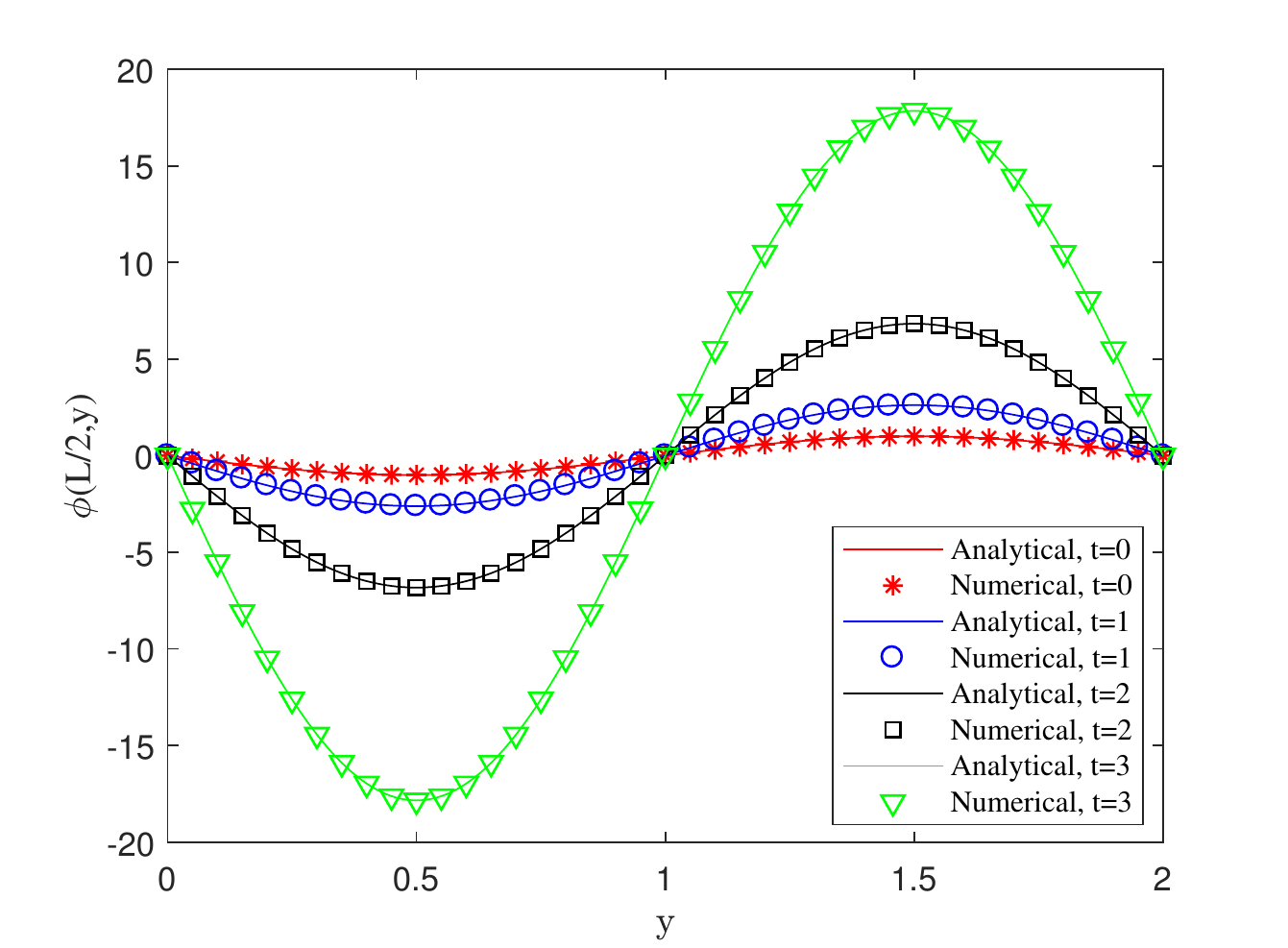}
    }
    \subfigure[]
    {
      \label{Fig4_1_1_b}
      \includegraphics[width=0.7\textwidth]{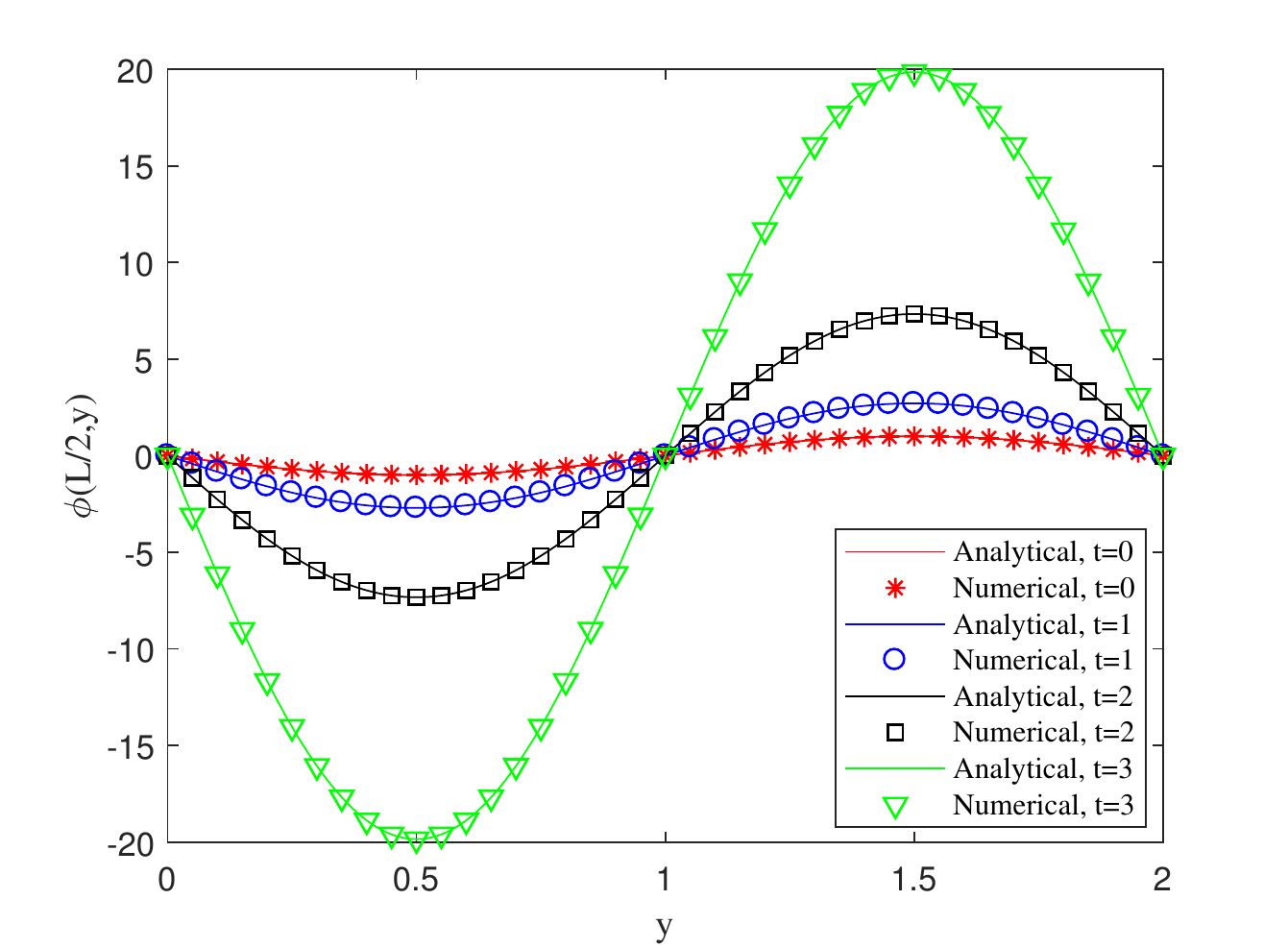}
    }
    \caption
    {
      Profiles of scalar variable $\phi$ at different P\'eclet numbers and time: (a) $Pe = 100$, (b) $Pe = 1000$.
    }
    \label{Fig4_1_1}
  \end{figure}

\begin{figure}[ht]
    \centering
    \subfigure[]
    {
      \label{Fig4_1_2_a}
      \includegraphics[width=0.7\textwidth]{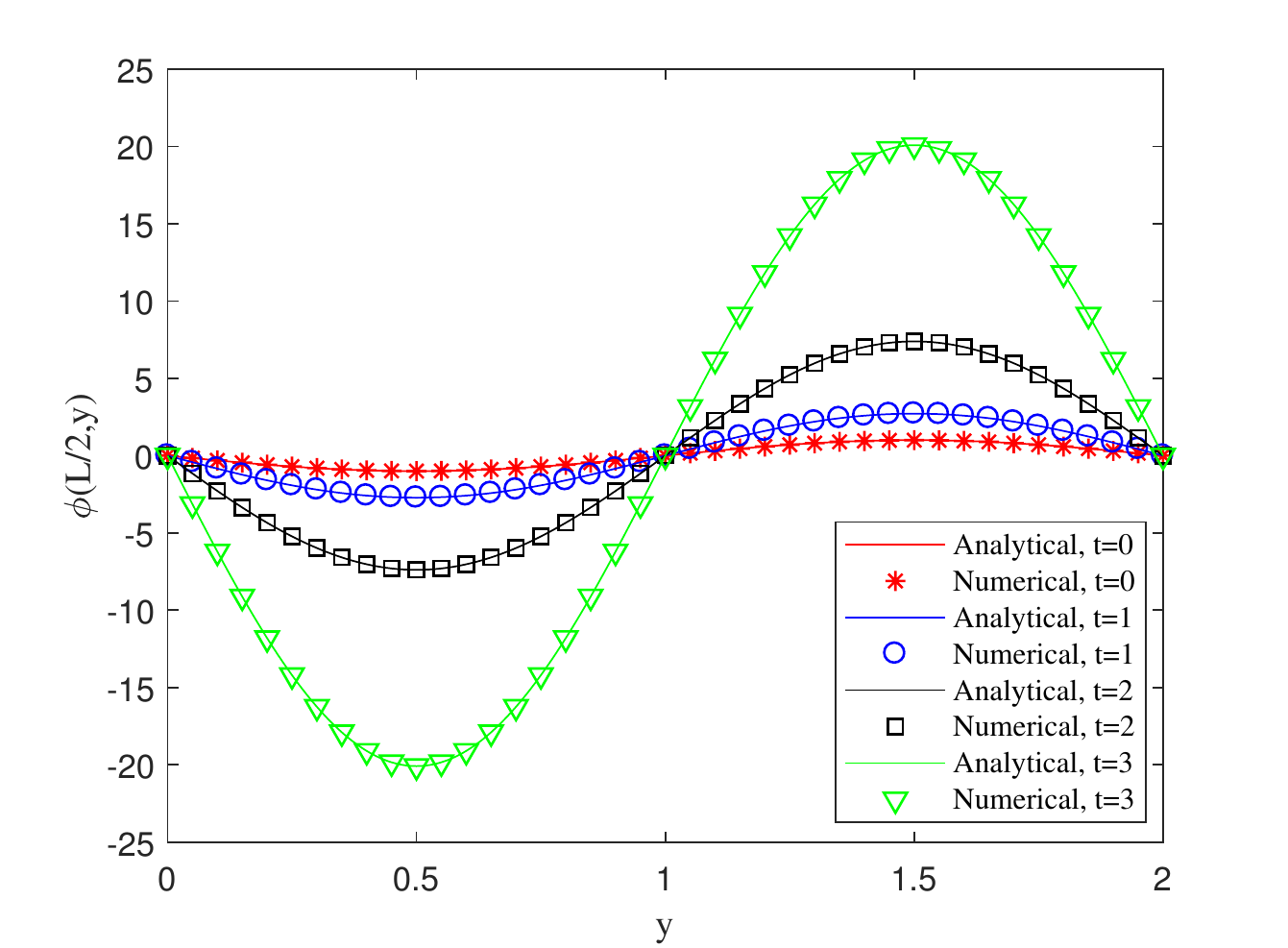}
    }
    \subfigure[]
    {
      \label{Fig4_1_2_b}
      \includegraphics[width=0.7\textwidth]{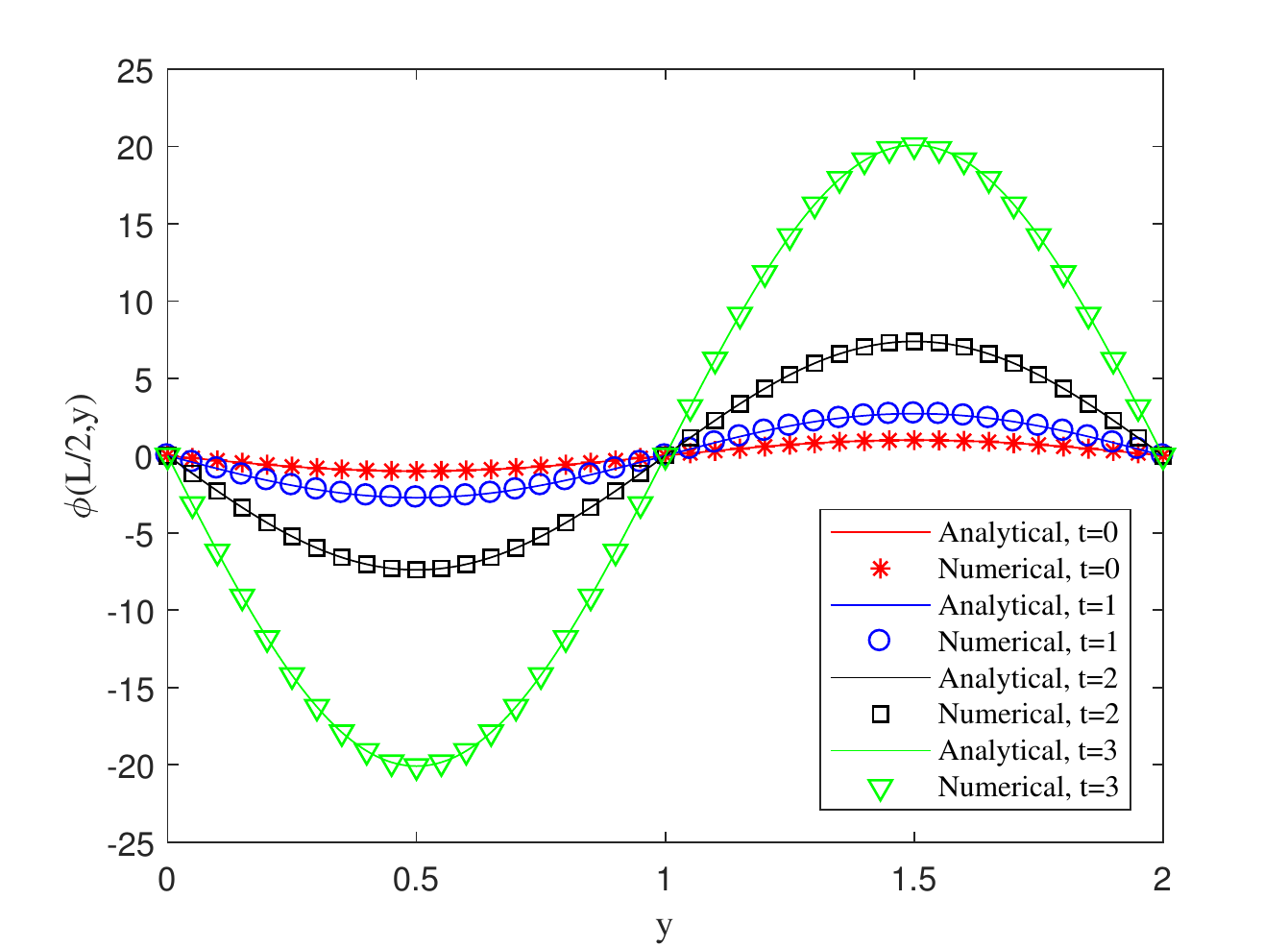}
    }
    \caption
    {
      Profiles of scalar variable $\phi$ at different P\'eclet numbers and time: (a) $Pe = 10^7$, (b) $Pe = 10^9$.
    }
    \label{Fig4_1_2}
  \end{figure}

  Now, we performed some simulations under different P\'eclet numbers and different time, where $Pe = L u_x / \alpha$, $L$ is the characteristic length (here $L = 2.0$), and the CFL condition number is equal to 0.5. The results are presented in Fig. \ref{Fig4_1_1} where $c = 1.0$, the uniform grid is $200 \times 200$, $\alpha$ can be determined by the specified $Pe$ (100 or 1000). As seen from the figure, the numerical solutions are in good agreement with analytical solutions. Besides, we also measured the values of $GRE$ at time $t = 3.0$, and they are $3.641 \times 10^{-4}$ for $Pe = 100$ and $4.109 \times 10^{-4}$ for $Pe = 1000$. In addition, to test the capacity of present DUGKS for this problem with a larger $Pe$, some simulations were performed with $Pe = 10^7$ and $10^9$, and the results are presented in Fig. \ref{Fig4_1_2}. From the Figure, we can find that the numerical solutions still agree well with the analytical solutions, and the values of $GRE$ at time $t = 3.0$ are $7.390 \times 10^{-5}$ for $Pe = 10^7$ and $7.383 \times 10^{-5}$ for $Pe = 10^9$. It is clearly that the deviations are small enough. We also performed a comparison among DUGKS, FV-LBM and MRT-LBM under the same conditions, and listed the results in Table \ref{table 4_1_1}. As we can see from this table, the performance of FV-LBM is worst, which is mainly caused by the severe numerical dissipation. Besides, the accuracies of DUGKS and MRT-LBM are almost the same.
  \begin{table}[ht]
    \caption{A comparison of DUGKS, FV-LBM and MRT-LBM}
    \label{table 4_1_1}
    \centering
    \begin{tabular} {lcccc}
      \hline\hline
                  &$Pe=100$              &$Pe=1000$               &$Pe=10^7$               &$Pe=10^9$      \\
      \midrule[1pt]
      DUGKS       &$3.641\times10^{-4}$ &$4.109\times10^{-4}$ &$7.390\times10^{-5}$ &$7.383\times10^{-5}$  \\
      FV-LBM      &$1.543\times10^{-3}$ &$1.431\times10^{-3}$ &$1.436\times10^{-3}$ &$1.436\times10^{-3}$  \\
      MRT-LBM     &$3.265\times10^{-4}$ &$1.709\times10^{-4}$ &$1.453\times10^{-4}$ &$1.452\times10^{-4}$  \\
      \hline\hline
    \end{tabular}
  \end{table}

\begin{figure}[ht]
    \centering
    \includegraphics[width=0.7\textwidth]{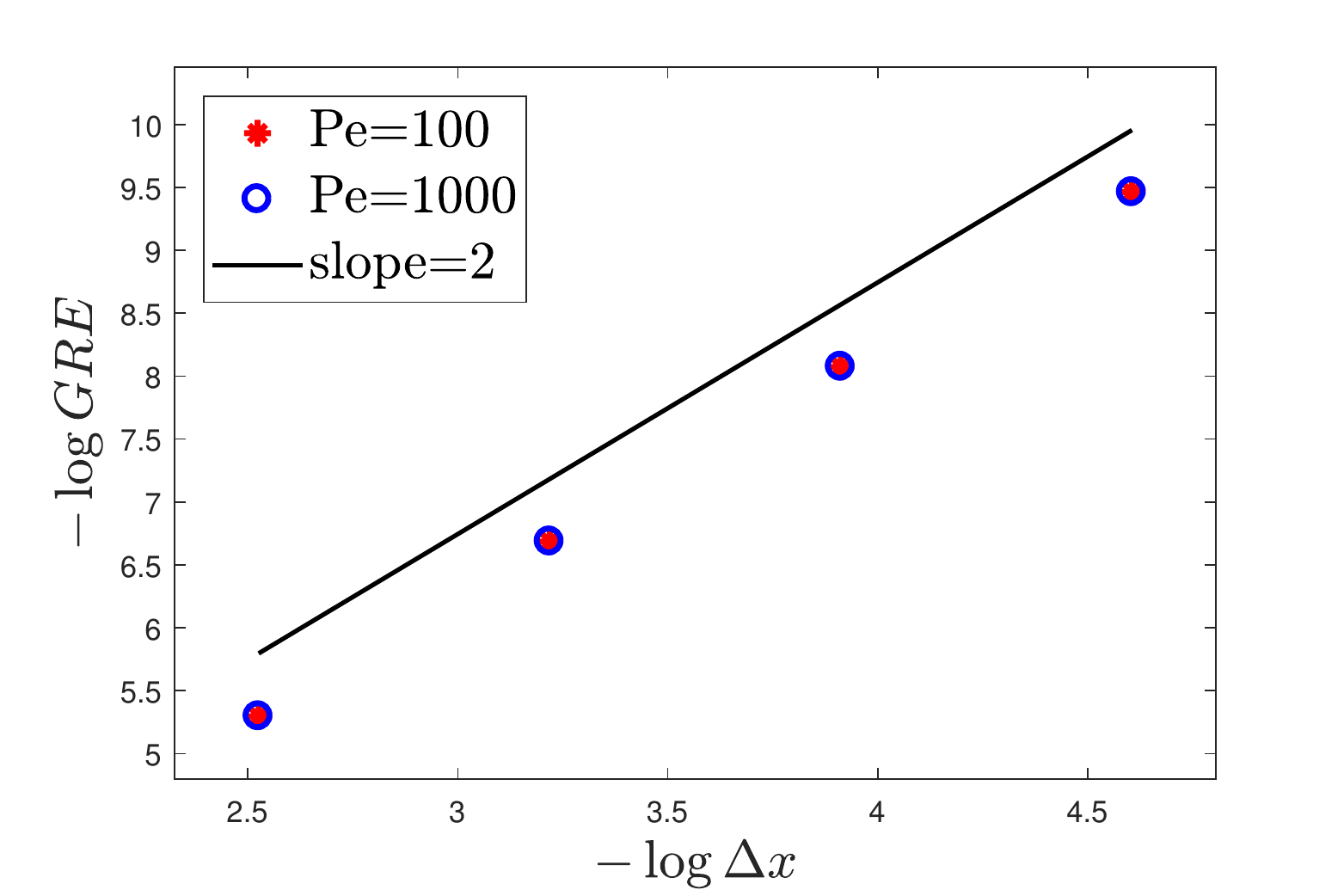}
    \caption
    {
      The global relative errors at different lattice sizes($\Delta x = L/25,L/50,L/100,L/200$), the slope of the $solid$ $line$ is 2.0, which indicates the present DUGKS has a second-order convergence rate in space.
    }
    \label{Fig4_1_3}
  \end{figure}

  Finally, the problem is applied to test the convergence rate of the present DUGKS. Since it is a periodic problem, the effect of the boundary conditions can be excluded. To this end, we carried out some simulations with different lattice sizes ($25 \times 25$ $\sim$ $200 \times 200$), and the time step is fixed at $1.0\times 10^{-5}$. As seen from Fig. \ref{Fig4_1_3}, that the present DUGKS has a second-order convergence rate.

  \textbf{Example 4.2} The Burgers-Fisher equation in two dimensions \cite{chai2016multiple} can be written as
  \begin{equation}
    \partial_t \phi + a\phi^\delta \partial_x \phi - b(\partial_{xx}\phi + \partial_{yy}\phi) - k\phi(1-\phi^\delta) = 0, \quad \delta > 1.
    \label{eq:4_2_1}
  \end{equation}
  The analytical solution of Eq. (\ref{eq:4_2_1}) can be given by \cite{chai2016multiple}
  \begin{equation}
    \phi(x,y,t) = \{\frac{1}{2}+ \frac{1}{2}\tanh[A(x+y-mt)]\}^{1/\delta},
    \label{eq:4_2_2}
  \end{equation}
  where $A=-\frac{a\delta}{4b(\delta+1)}$, $m=\frac{a^2+2bk(\delta+1)^2}{a(\delta+1)}$, $a$, $b$, $k$ and $\delta$ are constants. Different from the first problem, this problem is nonlinear, and boundary conditions are nonperiodic.
  
\begin{figure}
    \centering
    \includegraphics[width=0.7\textwidth]{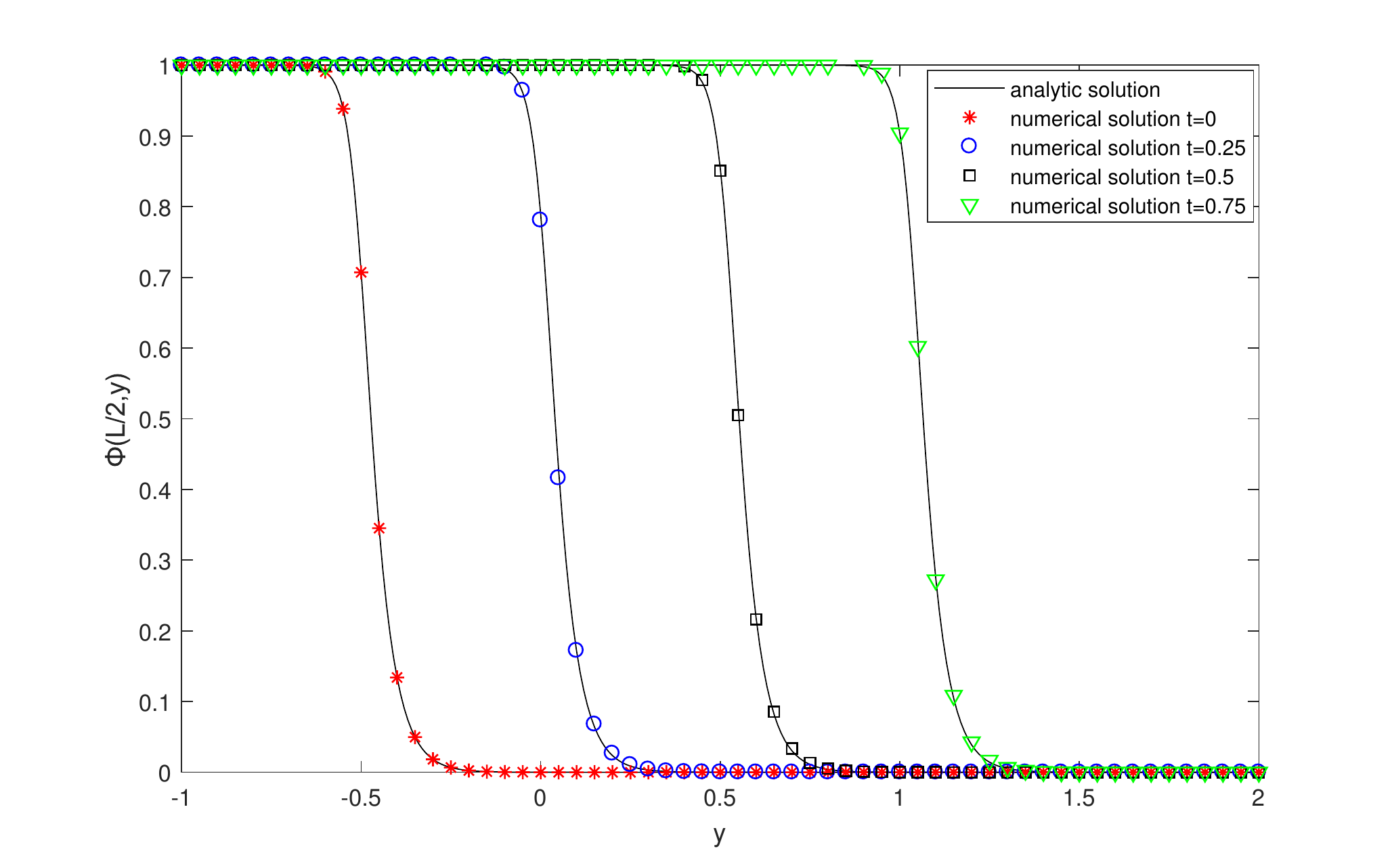}
    \caption{Profile of the scalar variable $\phi$ at different time.}
    \label{Fig4_2_1}
  \end{figure}

  For this problem, $\mathbf{B}=(\frac{a}{\delta+1},0)^T \phi^{\delta+1}$, and the simulations are performed on [-1,2] $\times$ [-1,2] with a 300 $\times$ 300 uniform grid size. In our simulations, parameters are set as $\delta=2.0$, $k=1.0$, $c=10$, $a=6.0$, $b=0.05$, and the CFL condition number is equal to 0.1. We presented the result in Fig. \ref{Fig4_2_1}, and found that the numerical solutions are in good agreement with the corresponding analytical solutions.
  
\begin{figure}
    \centering
    \includegraphics[width=0.7\textwidth]{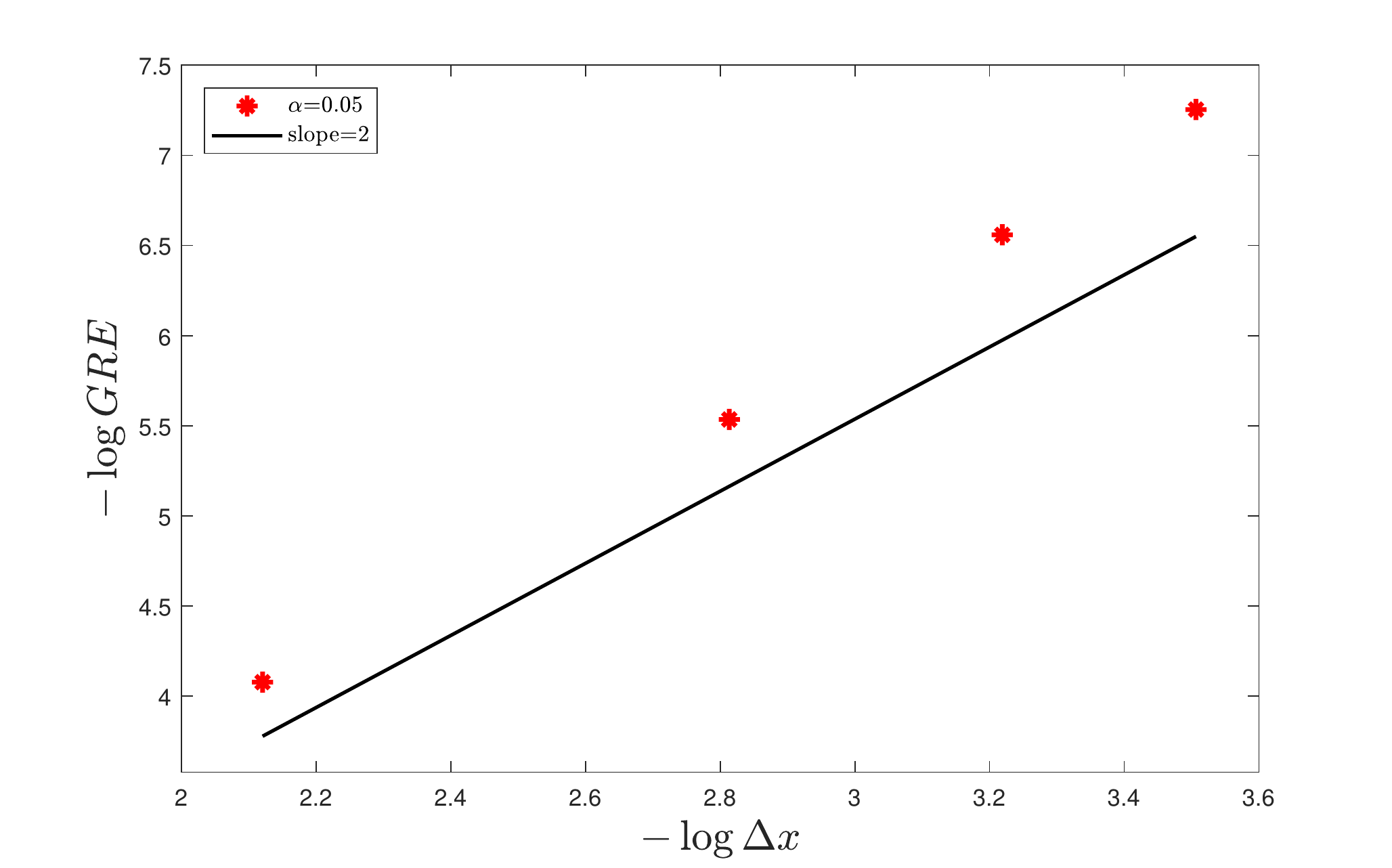}
    \caption{The global relative errors at different lattice sizes($\Delta x=L/25,L/50,L/75,L/100$), the slope of the $solid line$ is $2.0$, which indicates the present DUGKS has a second-order convergence rate in space.}
    \label{Fig4_2_2}
  \end{figure}

  To test the convergence rate of DUGKS for this problem, some simulations were carried out at time $t=1.0$, the lattice sizes are varied from $25 \times 25$ to $100 \times 100$, and time step $\Delta t=1.0 \times 10^{-5}$. From the results in Fig. \ref{Fig4_2_2}, it is clearly that the present DUGKS has a second-order convergence rate in space.

  As a finite-volume scheme, the DUGKS has the distinct advantage in adopting the non-uniform mesh. To show the advantage more clearly, we also performed some simulations on rectangular grid (lattice size is $300$ $\times$ $150$), and the other parameters are the same as above. We presented a comparison between uniform and non-uniform grids in Table \ref{table 4_2_1} where time $t=0.5$. As seen from this table, the errors of DUGKS with rectangular grid and those with uniform grid are of the same accuracy. While, computational cost of DUGKS with uniform grid ($300 \times 300$) is about twice as that of DUGKS rectangular grid ($300 \times 150$), the model on rectangular grid is more efficient than that on uniform grid.
  \begin{table}[ht]
    \caption{$GRE$s of the DUGKS model with uniform grid (denoted as DUGKS$^a$) and rectangular grid (denoted as DUGKS$^b$)}
    \label{table 4_2_1}
    \centering
    \begin{tabular} {llcccc}
      \hline\hline
              &            &$b=0.05$              &$b=0.1$               &$b=0.5$               &$b=1.0$                \\
      \midrule[1pt]
      $c=10$  &DUGKS$^a$   &$1.0826\times10^{-3}$ &$1.8261\times10^{-3}$ &$3.5657\times10^{-3}$ &$2.2113\times10^{-3}$  \\
              &DUGKS$^b$   &$1.1406\times10^{-3}$ &$1.8489\times10^{-3}$ &$3.6188\times10^{-3}$ &$2.1893\times10^{-3}$  \\
      $c=20$  &DUGKS$^a$   &$4.0612\times10^{-4}$ &$5.3645\times10^{-4}$ &$1.0287\times10^{-3}$ &$6.9190\times10^{-4}$  \\
              &DUGKS$^b$   &$5.9501\times10^{-4}$ &$5.8983\times10^{-4}$ &$1.0694\times10^{-3}$ &$7.3166\times10^{-4}$  \\
      \hline\hline
    \end{tabular} 
  \end{table}

  \textbf{Example 4.3} The generalized two dimensions NHCE in \cite{shi2009lattice}
  \begin{equation}
    \phi_t - \alpha(\phi^\delta)_{xx} - \alpha(\phi^\delta)_{yy} -\delta + \phi^\delta =0, \quad \delta>1,
    \label{eq:4_3_1}
  \end{equation}
  has the following analytical solution,
  \begin{equation}
    \phi(x,y,t) = \left\{\frac{1}{2} - \frac{1}{2}\tanh[\frac{\delta-1}{2\delta\sqrt{2\alpha}t}]\right\}^{-1/(\delta-1)},
    \label{eq:4_3_2}
  \end{equation}
  where $\alpha$ and $\delta$ are constants.

  For this problem, we take $\mathbf{B}=0$ and $\mathbf{D}=\phi^\delta \mathbf{I}$, which leads to the following equilibrium distribution function
  \begin{equation}
    f_i^{eq}=\omega_i\left[\phi + \frac{(\mathbf{D}- \phi\mathbf{I}):(\mathbf{c}_i\mathbf{c}_i-c_s^2\mathbf{I})}{2c_s^2}\right].
    \label{eq:4_3_3}
  \end{equation}

\begin{figure}[ht]
    \centering
    \subfigure[]
    {
      \label{Fig4_3_1_a}
      \includegraphics[scale=0.4]{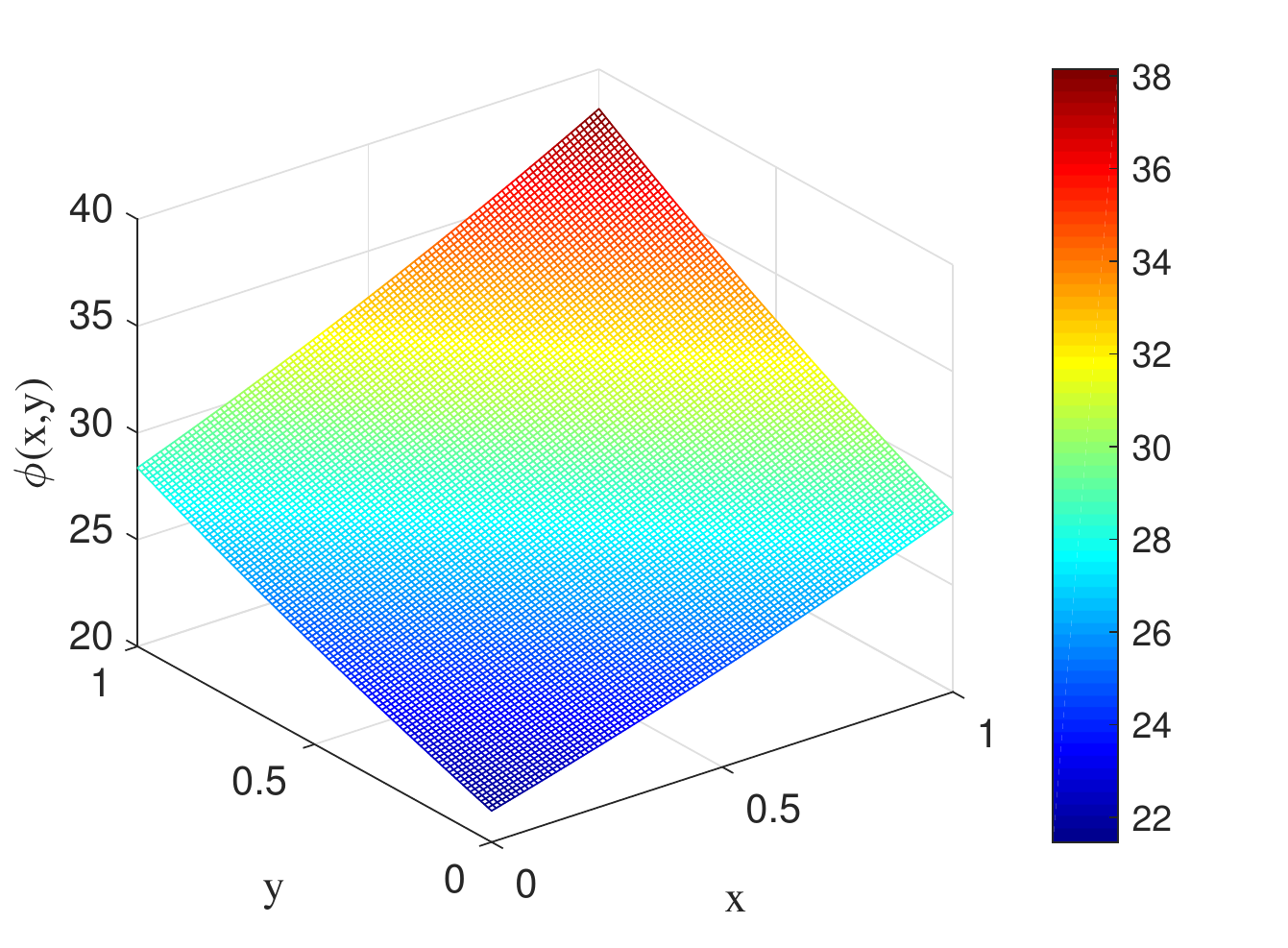}
    }
    \subfigure[]
    {
      \label{Fig4_3_1_b}
      \includegraphics[scale=0.4]{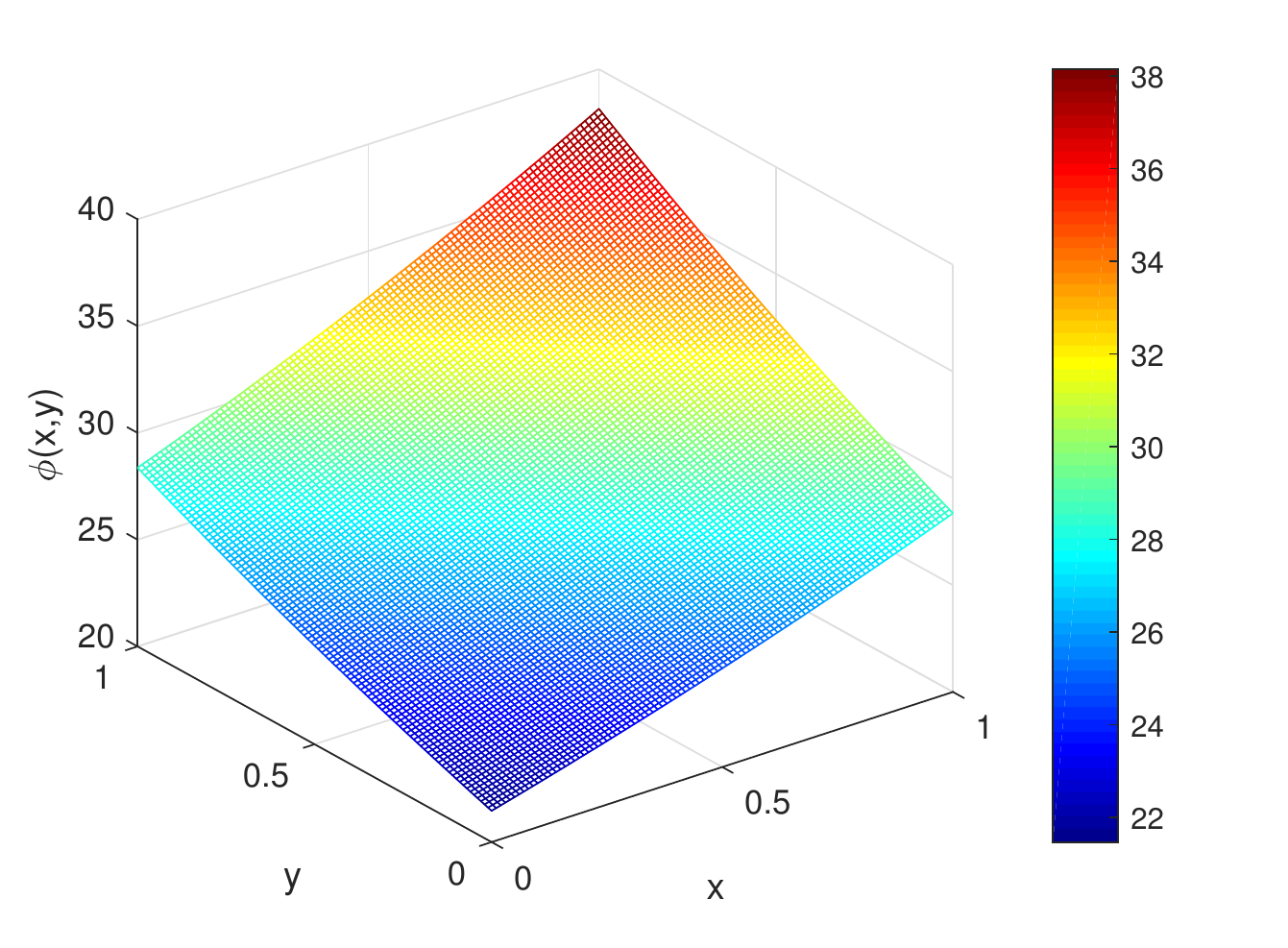}
    }
    \caption{Scalar variable $\phi$ at $t=1.0$, $\alpha=1.0$, $\delta=1.2$: (a) numerical solution, (b) analytical solution.}
    \label{Fig4_3_1}
  \end{figure}

\begin{figure}[ht]
    \centering
    \subfigure[]
    {
      \label{Fig4_3_2_a}
      \includegraphics[scale=0.4]{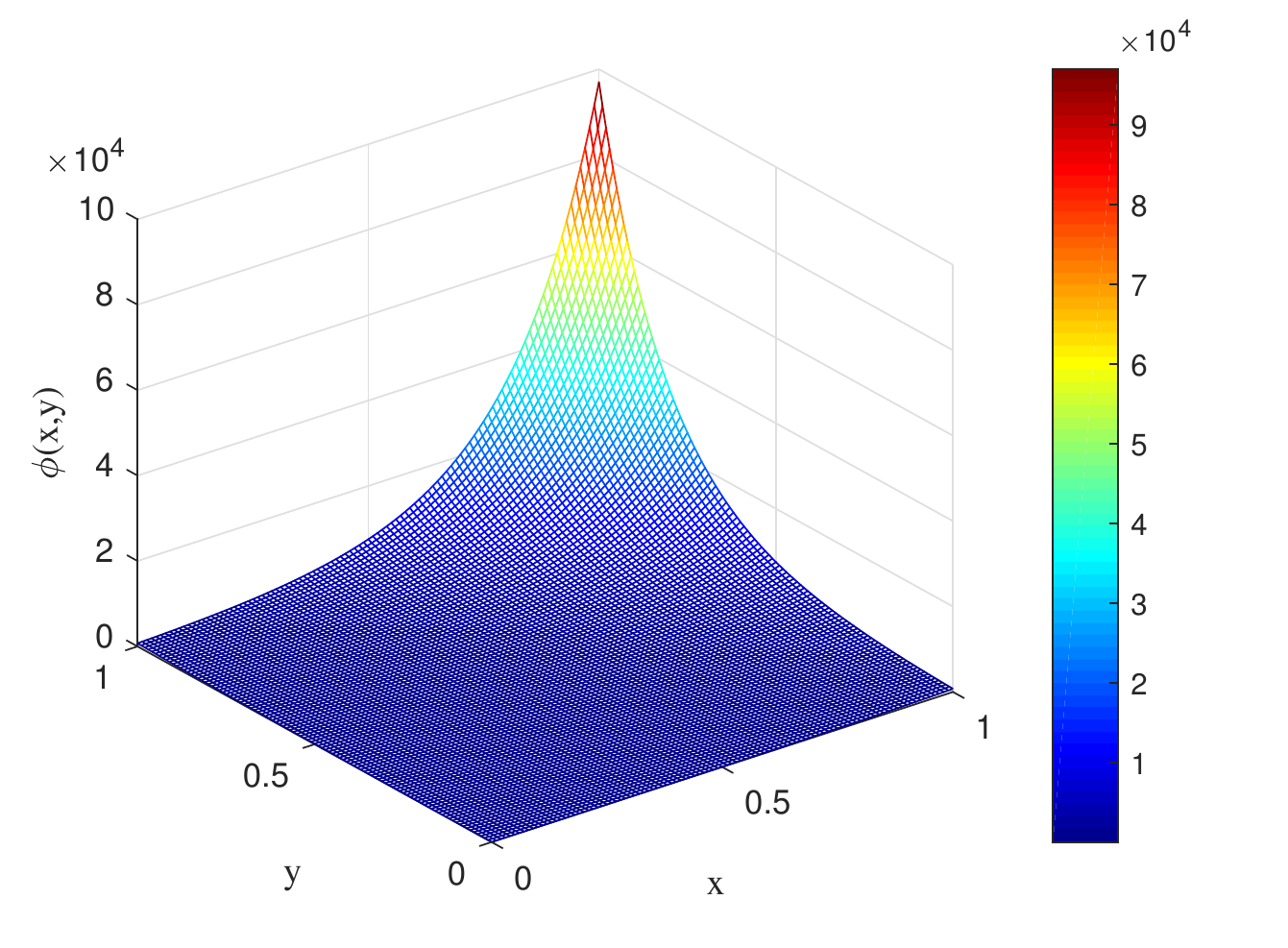}
    }
    \subfigure[]
    {
      \label{Fig4_3_2_b}
      \includegraphics[scale=0.4]{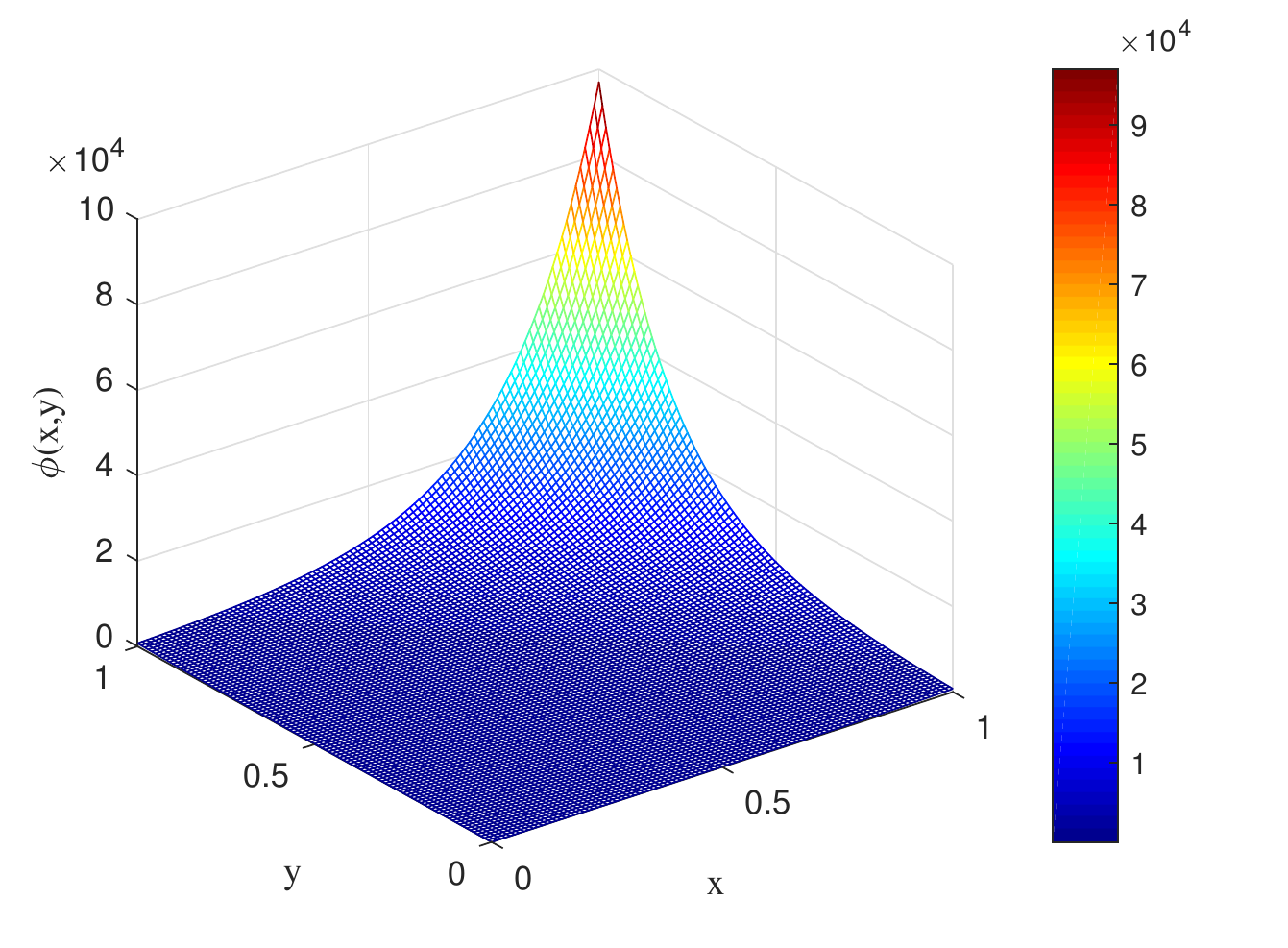}
    }
    \caption{Scalar variable $\phi$ at $t=1.0$, $\alpha=0.01$, $\delta=1.2$: (a) numerical solution, (b) analytical solution.}
    \label{Fig4_3_2}
  \end{figure}

\begin{figure}[ht]
    \centering
    \includegraphics[scale=0.7]{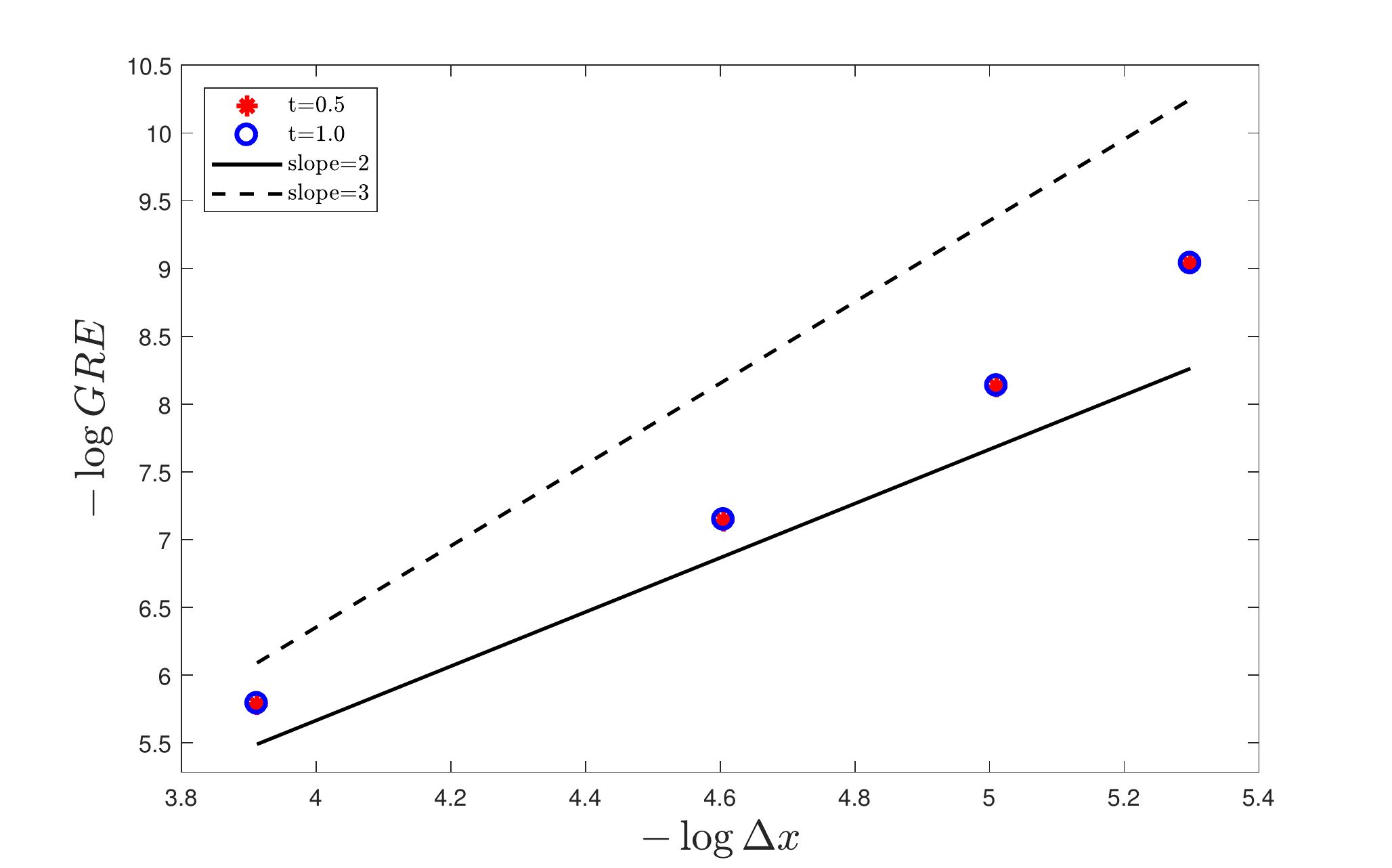}
    \caption{The global relative errors at different lattice sizes($\Delta x=L/50,L/100,L/150,L/200$), the slope of the $solid line$ is $2.0$ and the slope of the $dash line$ is $3.0$, which indicates the present DUGKS model has a second-order convergence rate in space.}
    \label{Fig4_3_3}
  \end{figure}

  We carried out some simulations on $[0,1]\times[0,1]$ with the lattice size $100 \times 100$. As seen from Fig. \ref{Fig4_3_1} and \ref{Fig4_3_2}, the numerical solutions are close to the analytical solutions at different values of $\alpha$, and the gradient term $\nabla \phi$ increases very fast with the decrease of $\alpha$. To see the difference between analytical and numerical solutions, we also measured the global relative errors and present them in Table \ref{table 4.3.1} with CFL condition number equaling to 0.1. To test the convergence rate of DUGKS for this problem, We plotted the global relative errors at different lattice size in Fig. \ref{Fig4_3_3} where $dt=1.0\times 10^{-6}$, $\delta=1.2$ and $\alpha=0.01$. From this figure, it is also found that the DUGKS model for the NHCE is of second-order accuracy in space.
  \begin{table}[ht]
    \caption{The global relative errors with different values of $\alpha$ and $c$ at $t=1.0$ and $\delta=1.2$}
    \label{table 4.3.1}\centering
    \begin{tabular}{lcccc}
      \hline\hline
                & $\alpha=0.01$          & $\alpha=0.05$          & $\alpha=0.1$           & $\alpha=1.0$          \\
      \midrule[1pt]
      $c=10 $   & $1.4858\times10^{-3}$  & $4.5385\times10^{-4}$  & $4.3775\times10^{-4}$  & $3.9429\times10^{-4}$ \\
      $c=20 $   & $1.0398\times10^{-3}$  & $2.5046\times10^{-4}$  & $1.3507\times10^{-4}$  & $1.0327\times10^{-4}$ \\
      $c=100$   & $9.9290\times10^{-5}$  & $1.5214\times10^{-4}$  & $1.0280\times10^{-4}$  & $2.5874\times10^{-5}$ \\
      \hline\hline
    \end{tabular}
  \end{table}

\begin{figure}[ht]
    \subfigure[]
    {
      \label{Fig4_3_4_a}
      \includegraphics[scale=0.5]{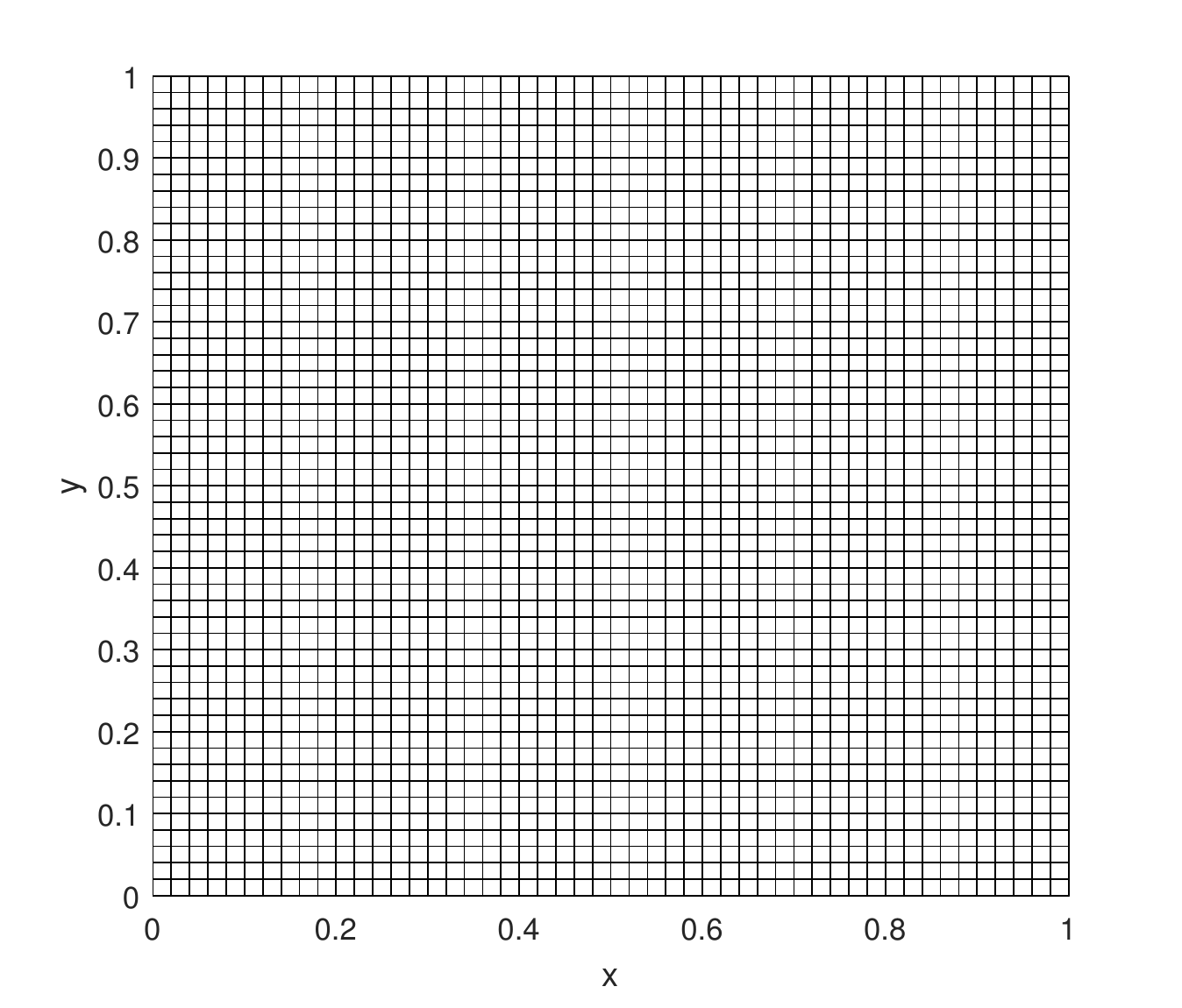}
    }
    \subfigure[]
    {
      \label{Fig4_3_4_b}
      \includegraphics[scale=0.5]{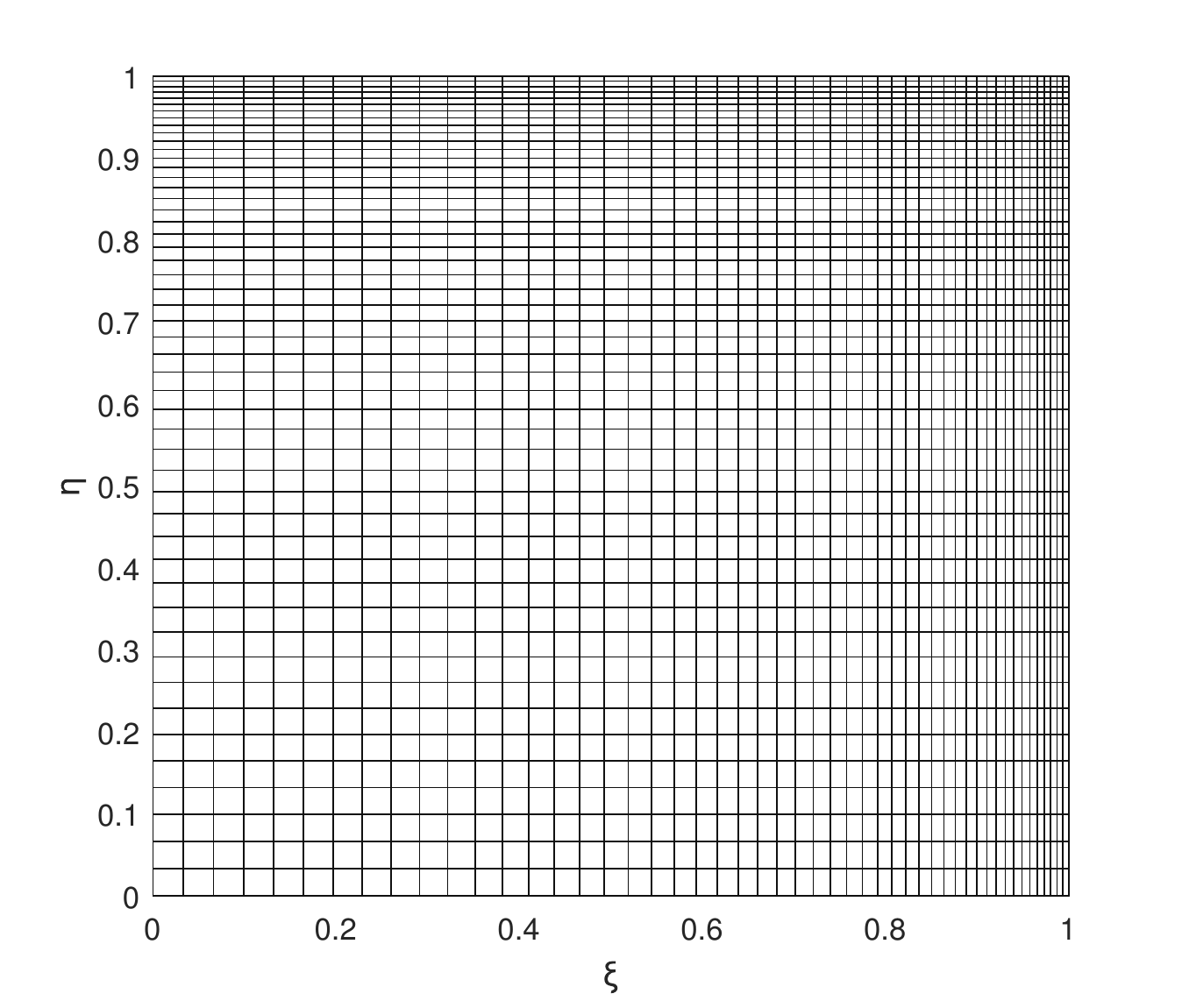}
    }
    \caption{Distributions of uniform and non-uniform meshes}
    \label{Fig4_3_4}
  \end{figure}

  For this problem, some simulations were also performed with the non-uniform mesh. The non-uniform mesh is generated by the following transformation,
  \begin{equation}
    x=\frac{\tanh(k\eta)}{\tanh(k)},
    \label{eq:4_3_4}
  \end{equation}
  \begin{equation}
    y=\frac{\tanh(k\zeta)}{\tanh(k)},
    \label{eq:4_3_5}
  \end{equation}
  where $k=1.5$, which is used to control the distribution of non-uniform mesh. The grid point in ($\xi, \eta$) plane are defined by $\xi_i=i/Nx$ and $\eta_j = j/Ny$ for $i=0,1,...,N_x$ and $j=0,1,...,N_y$. The distributions of the uniform and the non-uniform meshes used in our simulations are shown in Fig. \ref{Fig4_3_4}. With the same parameters shown in Table \ref{table 4.3.1}, we carried out some simulations with the non-uniform mesh, and the results are listed in Table \ref{table 4_3_2}.
  \begin{table}[ht]
    \caption{The global relative errors at $t=1.0$, $\delta=1.2$, $c=10.0$ and different values of $\alpha$ and meshes (mesh$^a$ denotes the uniform mesh with the lattice size $100\times100$, mesh$^b$ denotes the non-uniform mesh with the lattice size $100\times100$, mesh$^c$ denotes the non-uniform mesh with the lattice size $50\times50$)}
    \label{table 4_3_2}
    \centering
    \begin{tabular}{lcccc}
      \hline\hline
                & $\alpha=0.01$          & $\alpha=0.05$          & $\alpha=0.1$           & $\alpha=1.0$          \\
      \midrule[1pt]
      mesh$^a$  & $1.4858\times10^{-3}$  & $4.5385\times10^{-4}$  & $4.3775\times10^{-4}$  & $3.9429\times10^{-4}$ \\
      mesh$^b$  & $3.3793\times10^{-4}$  & $3.2038\times10^{-4}$  & $3.3250\times10^{-4}$  & $3.6957\times10^{-4}$ \\
      mesh$^c$  & $1.0070\times10^{-3}$  & $4.1587\times10^{-4}$  & $3.5491\times10^{-4}$  & $3.7804\times10^{-5}$ \\
      \hline\hline
    \end{tabular}
  \end{table}

  As seen from Table \ref{table 4_3_2}, the $GRE$s with non-uniform meshes are smaller than those of mesh$^a$, and the difference becomes more obvious with the decrease of $\alpha$. This illustrates that the appropriate non-uniform mesh can improve the accuracy of the present DUGKS.
 
  \textbf{Example 4.4} The Gaussian hill problem is described by the following anisotropic convection diffusion equation \cite{chai2016multiple}
  \begin{equation}
    \partial_t\phi + \nabla \cdot (\phi\mathbf{u}) = \nabla \cdot (\mathbf{K} \cdot \nabla \phi),
    \label{eq:4_4_1}
  \end{equation}
  where $\mathbf{u}=(u_x, u_y)^T$ is a constant velocity, $\mathbf{K}$ is the constant diffusion tensor, and can be defined as
  \begin{equation}
    \mathbf{K}=
    \left(
      \begin{array}{cc}
        \kappa_{xx} & \kappa_{xy}\\
        \kappa_{yx} & \kappa_{yy}
      \end{array}
    \right).
    \label{eq:4_4_2}
  \end{equation}
  The analytical solution to this Gaussian hill problem can be expressed as
  \begin{equation}
    \phi(\mathbf{x},t)=\frac{\phi_0}{2\pi|\det(\sigma)|^{1/2}}\exp\left\{-\frac{\sigma^{-1}:[(\mathbf{x}-\mathbf{u}t)(\mathbf{x}-\mathbf{u}t)]}
    {2}\right\},
    \label{eq:4_4_3}
  \end{equation}
  where $\mathbf{x}=(x,y)^T$, $\sigma=\sigma_0^2\mathbf{I}+2\mathbf{K}t$, $\sigma^{-1}$ is the inverse matrix of $\sigma$, $\det(\sigma)$ is the determinant of $\sigma$.

  To study the Gaussian hill problem, we first write Eq. (\ref{eq:4_4_1}) in an isotropic form,
  \begin{equation}
    \partial_t\phi + \nabla\cdot(\phi\mathbf{u}) = \nabla\cdot[\kappa(\nabla\cdot\mathbf{D})],
    \label{eq:4_4_4}
  \end{equation}
  where $\mathbf{B} = \phi \mathbf{u}$ and the tensor $\mathbf{D}$ is given by $\mathbf{D}=\mathbf{K}\phi/\kappa$ with $\kappa$ being a positive constant. The physical domain of the problem $[-1,1] \times [-1,1]$ and the periodic boundary conditions are applied to all directions. In our simulations, $\sigma= 0.01$, $\mathbf{u}=(0.01,0.01)^T$, $\phi_0 = 2\pi\sigma_0^2$, CFL = 0.5 and the lattice size is $400\times400$. To test the capacity of the present DUGKS for the Gaussian hill problem, the following three types of diffusion tensor are considered,
  \begin{equation}
    \mathbf{K}=\left[
      \left(
        \begin{array}{cc}
          1 & 0\\
          0 & 1
        \end{array}
      \right),
      \left(
        \begin{array}{cc}
          1 & 0\\
          0 & 2
        \end{array}
      \right),
      \left(
        \begin{array}{cc}
          1 & 1\\
          1 & 2
        \end{array}
      \right)\right]
     \times 10^{-3},
    \label{eq:4_4_5}
  \end{equation}
  which are corresponding to the isotropic, diagonally anisotropic and fully anisotropic diffusion problems.

\begin{figure}[ht]
    \subfigure[]
    {
      \label{Fig4_4_1_a}
      \includegraphics[scale=0.5]{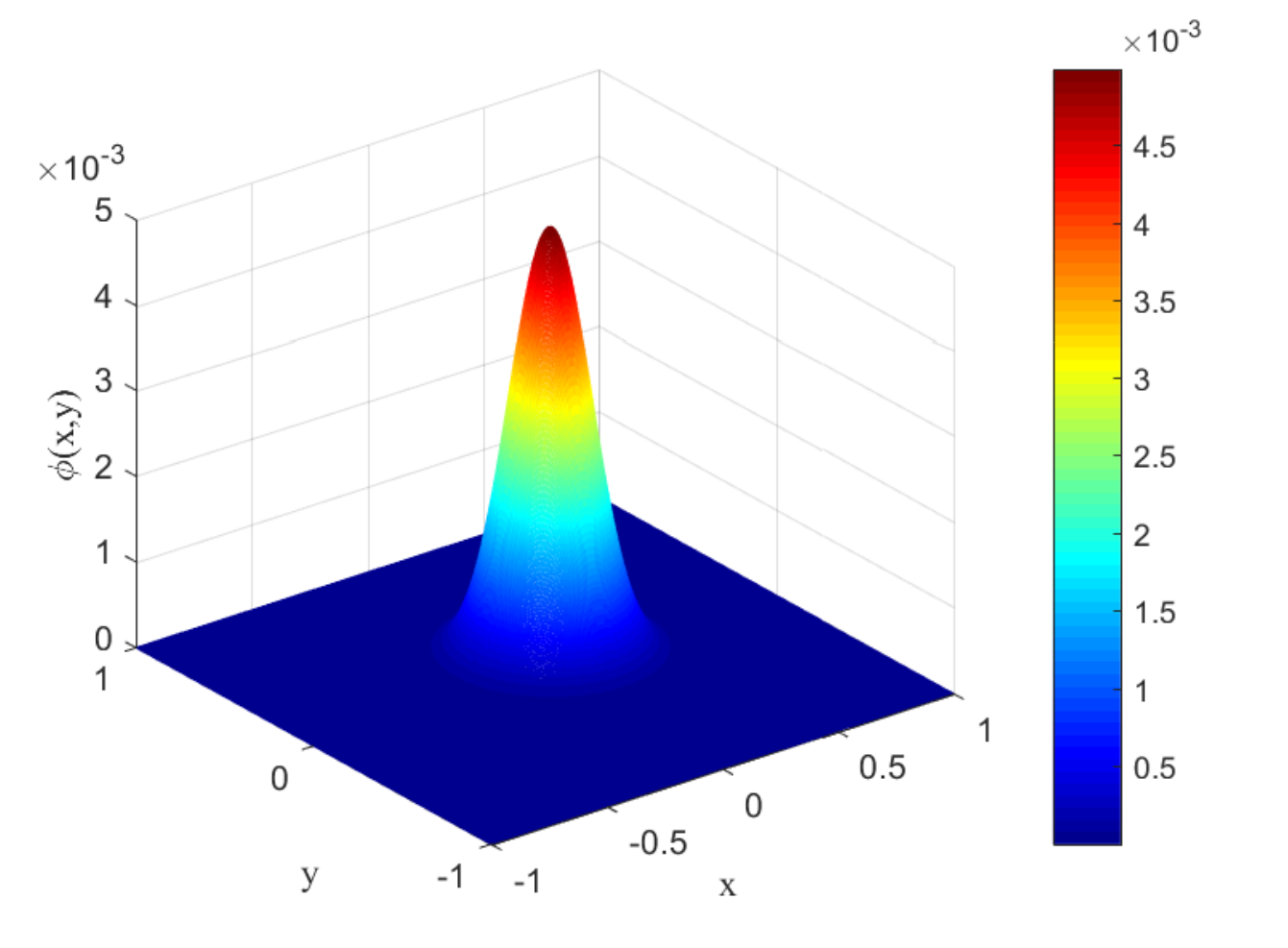}
    }
    \subfigure[]
    {
      \label{Fig4_4_1_b}
      \includegraphics[scale=0.5]{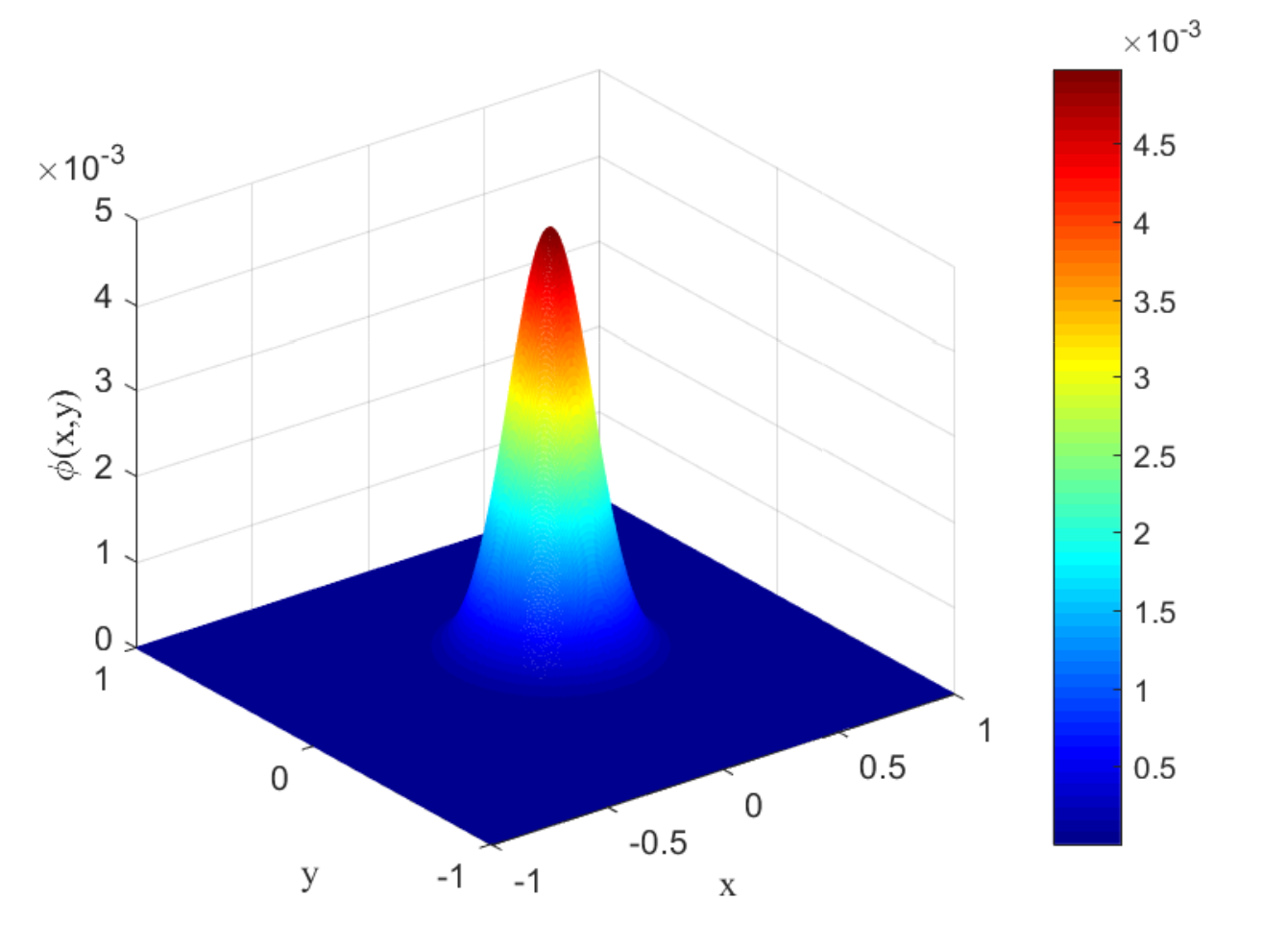}
    }
    \caption{Distributions of the scalar variable $\phi$ at time $t=10$ [isotropic diffusion problem: (a) numerical solution, (b) analytical solution]}
    \label{Fig4_4_1}
  \end{figure}

\begin{figure}[ht]
    \subfigure[]
    {
      \includegraphics[scale=0.5]{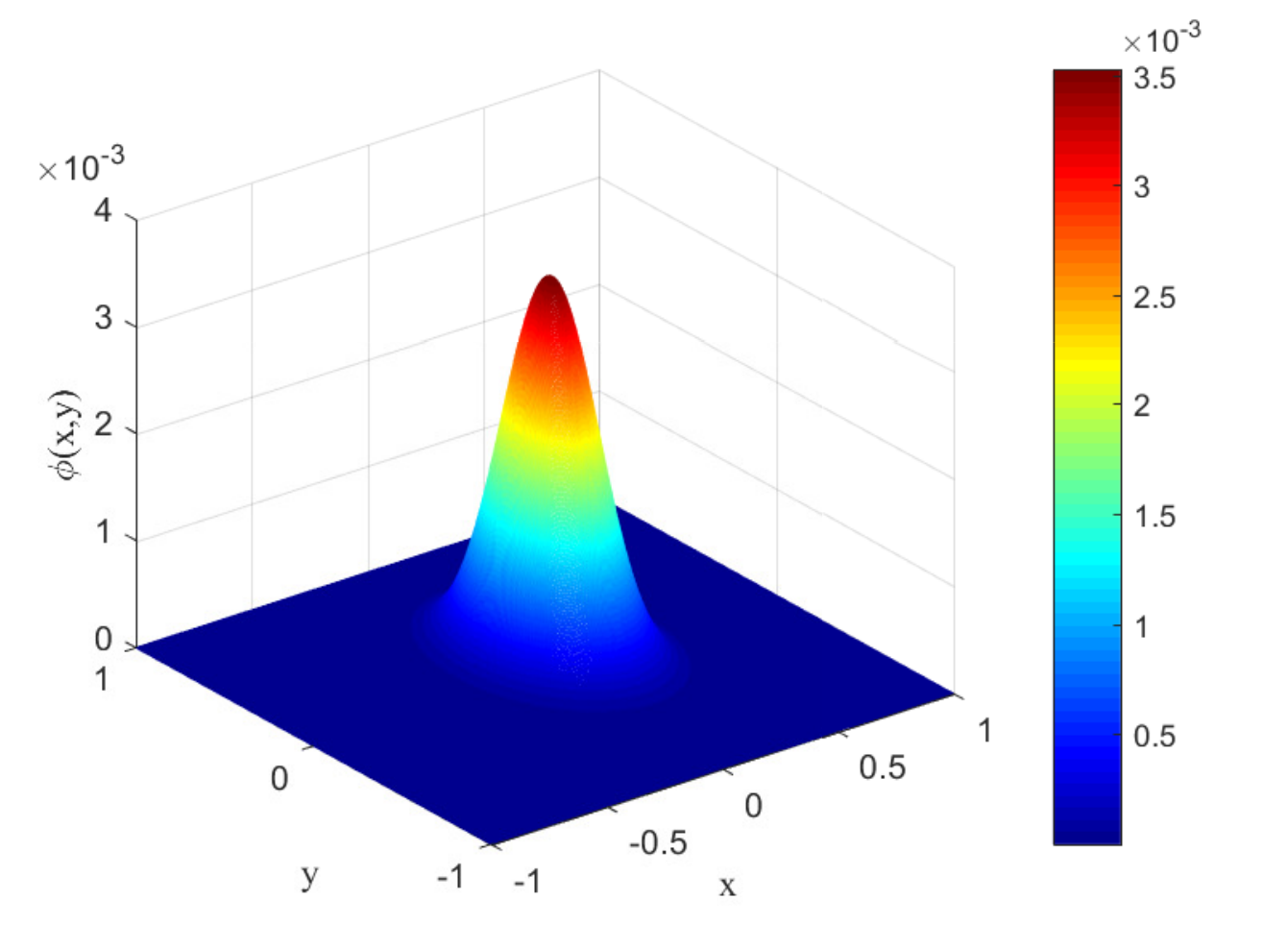}
      \label{Fig4_4_2_a}
    }
    \subfigure[]
    {
      \includegraphics[scale=0.5]{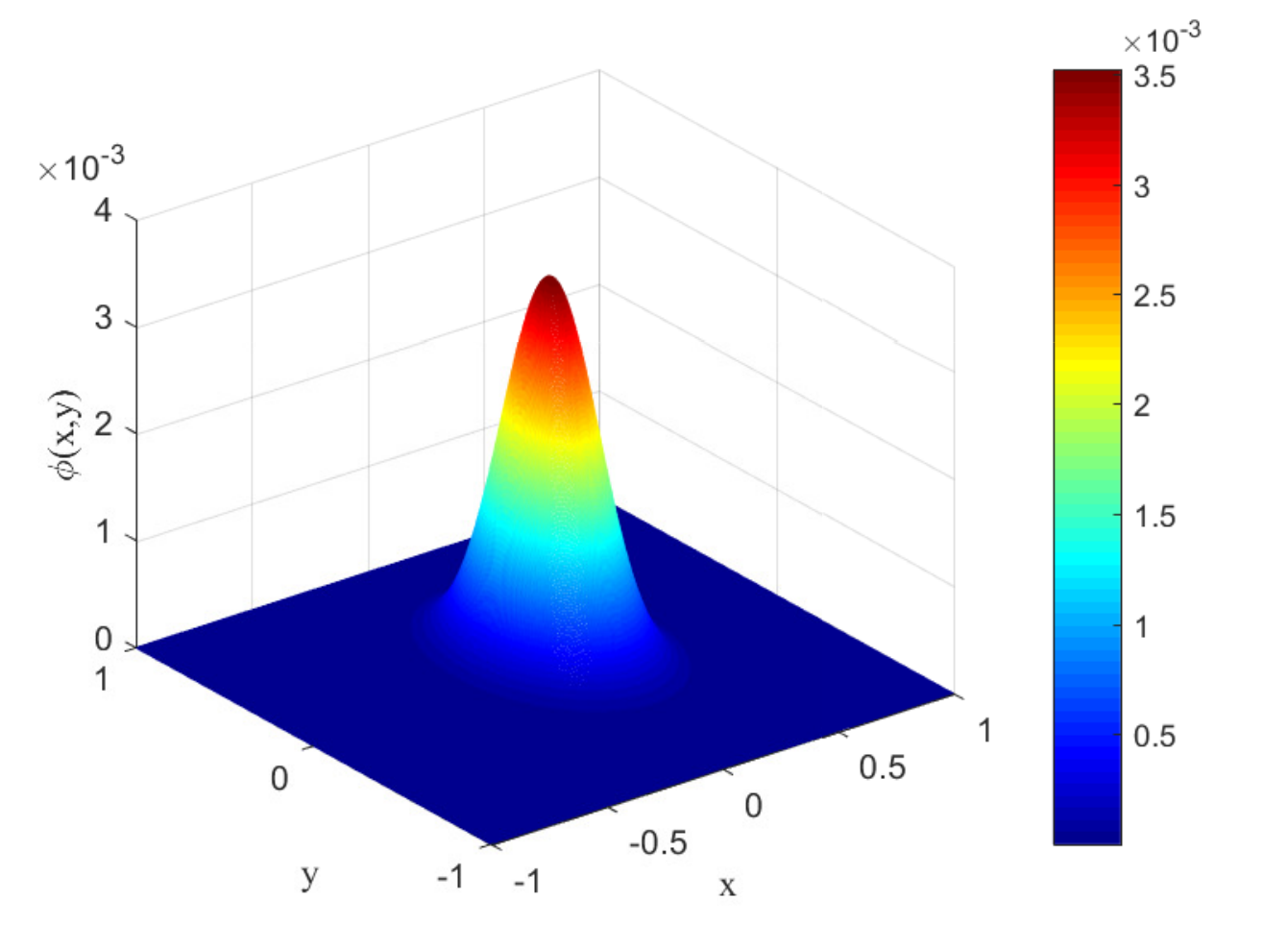}
      \label{Fig4_4_2_b}
    }
    \caption{Distributions of the scalar variable $\phi$ at time $t=10$ [diagonally anisotropic diffusion problem: (a) numerical solution, (b) analytical solution]}
    \label{Fig4_4_2}
  \end{figure}

\begin{figure}[ht]
    \subfigure[]
    {
      \includegraphics[scale=0.5]{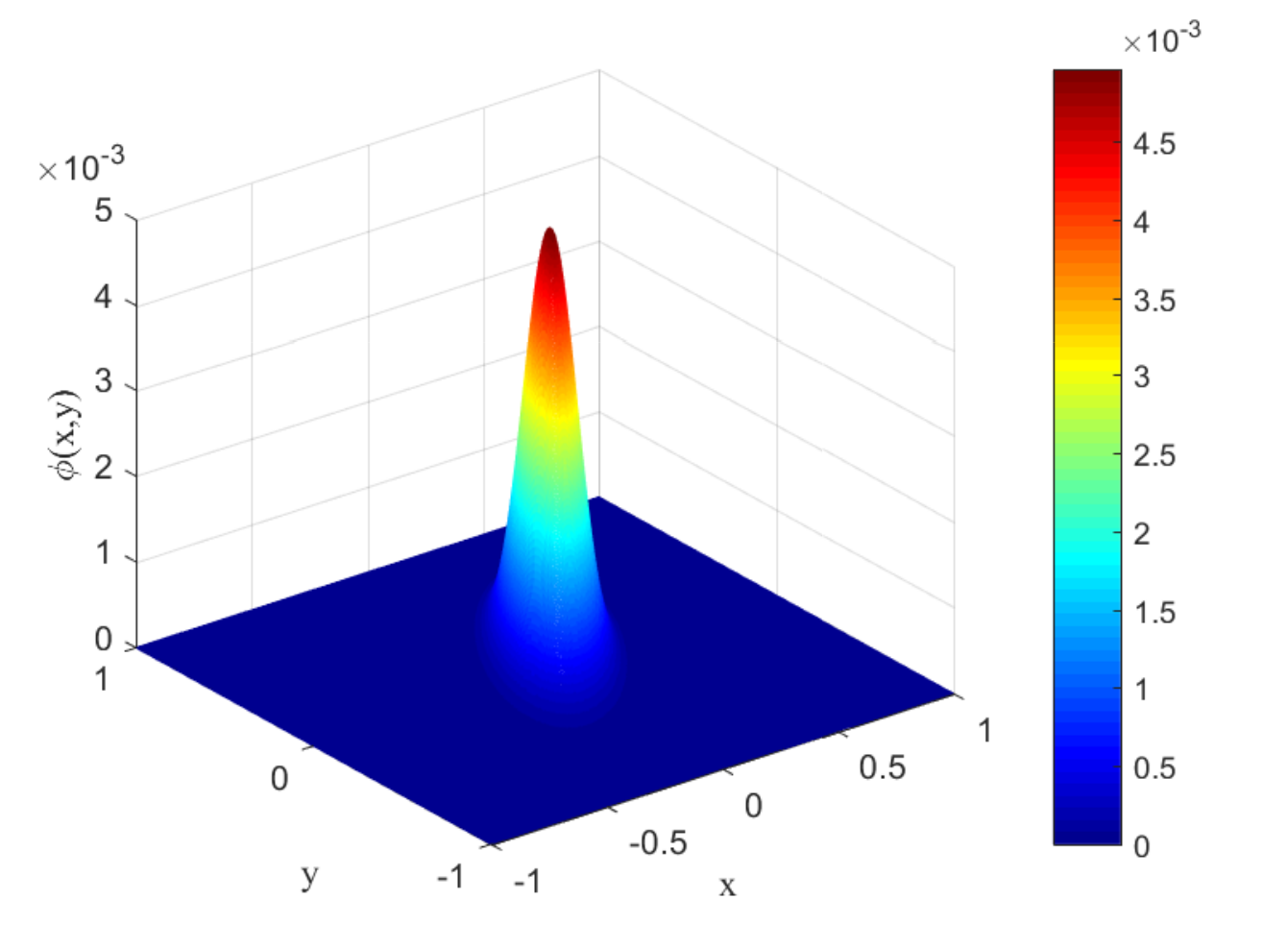}
      \label{Fig4_4_3_a}
    }
    \subfigure[]
    {
      \includegraphics[scale=0.5]{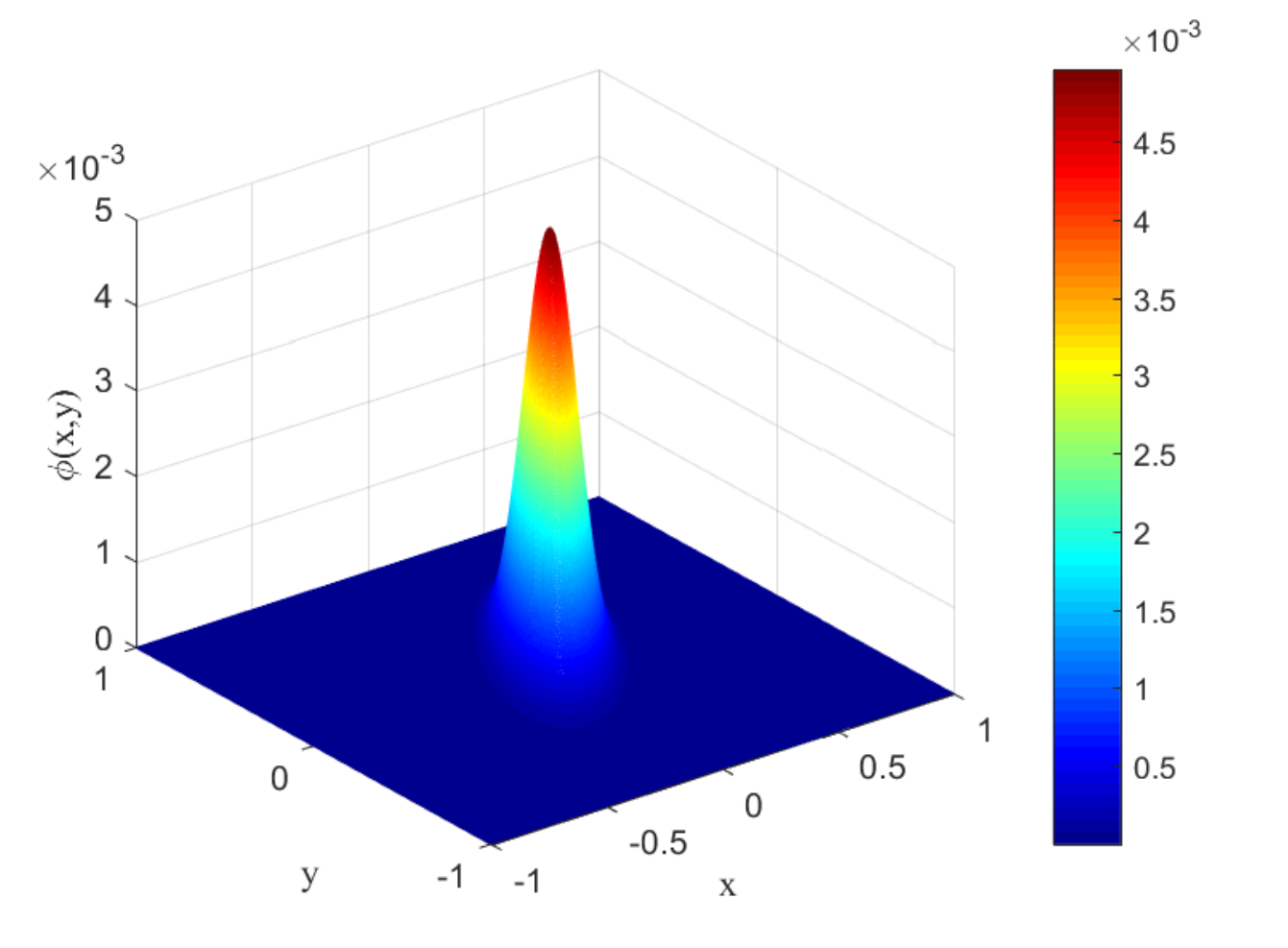}
      \label{Fig4_4_3_b}
    }
    \caption{Distributions of the scalar variable $\phi$ at time $t=10$ [fully anisotropic diffusion problem: (a) numerical solution, (b) analytical solution]}
    \label{Fig4_4_3}
  \end{figure}

\begin{figure}[ht]
    \centering
    \includegraphics[scale=0.7]{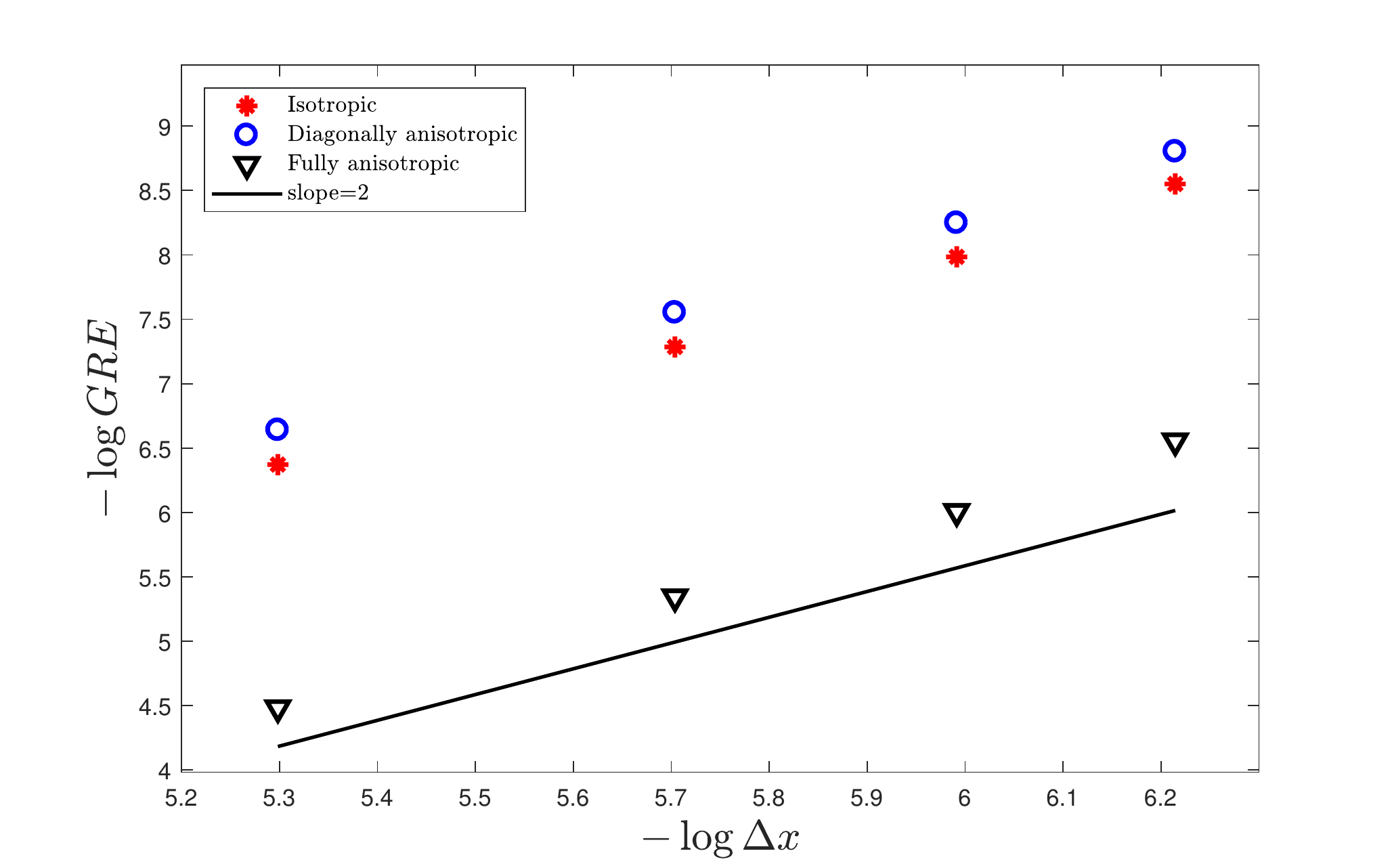}
    \caption{The global relative errors at different lattice sizes ($\Delta x=L/200, L/300, L/400, L/500$), the slope of the $solid line$ is 2.0, indicating present DUGKS has a second-order convergence rate in space.}
    \label{Fig4_4_4}
  \end{figure}

  We conducted several simulations and presented the numerical solutions at time $t=10$ in Figs. \ref{Fig4_4_1}, \ref{Fig4_4_2} and \ref{Fig4_4_3} where $\kappa=0.001$ and $c=1.0$. As shown in these figures, the numerical solutions are consistent with the analytical solutions. In addition, to see the deviation between the numerical and analytical solutions, the $GRE$s of isotropic, diagonally anisotropic and fully anisotropic diffusion problems are also calculated, and they are $1.0829 \times 10^{-3}$, $7.8158 \times 10^{-4}$, and $1.7746 \times 10^{-3}$, which also illustrate that the present DUGKS is accurate for Gaussian hill problem. Fig. \ref{Fig4_4_4} shows the accuracy of the present DUGKS for this problem, the lattice size is varied from $200 \times 200$ to $500 \times 500$ with time step $\Delta t=1.0\times 10^{-4}$. From this figure, we can find that the present DUGKS has a second-order convergence rate in space.

\begin{figure}[ht]
    \subfigure[]
    {
      \includegraphics[scale=0.5]{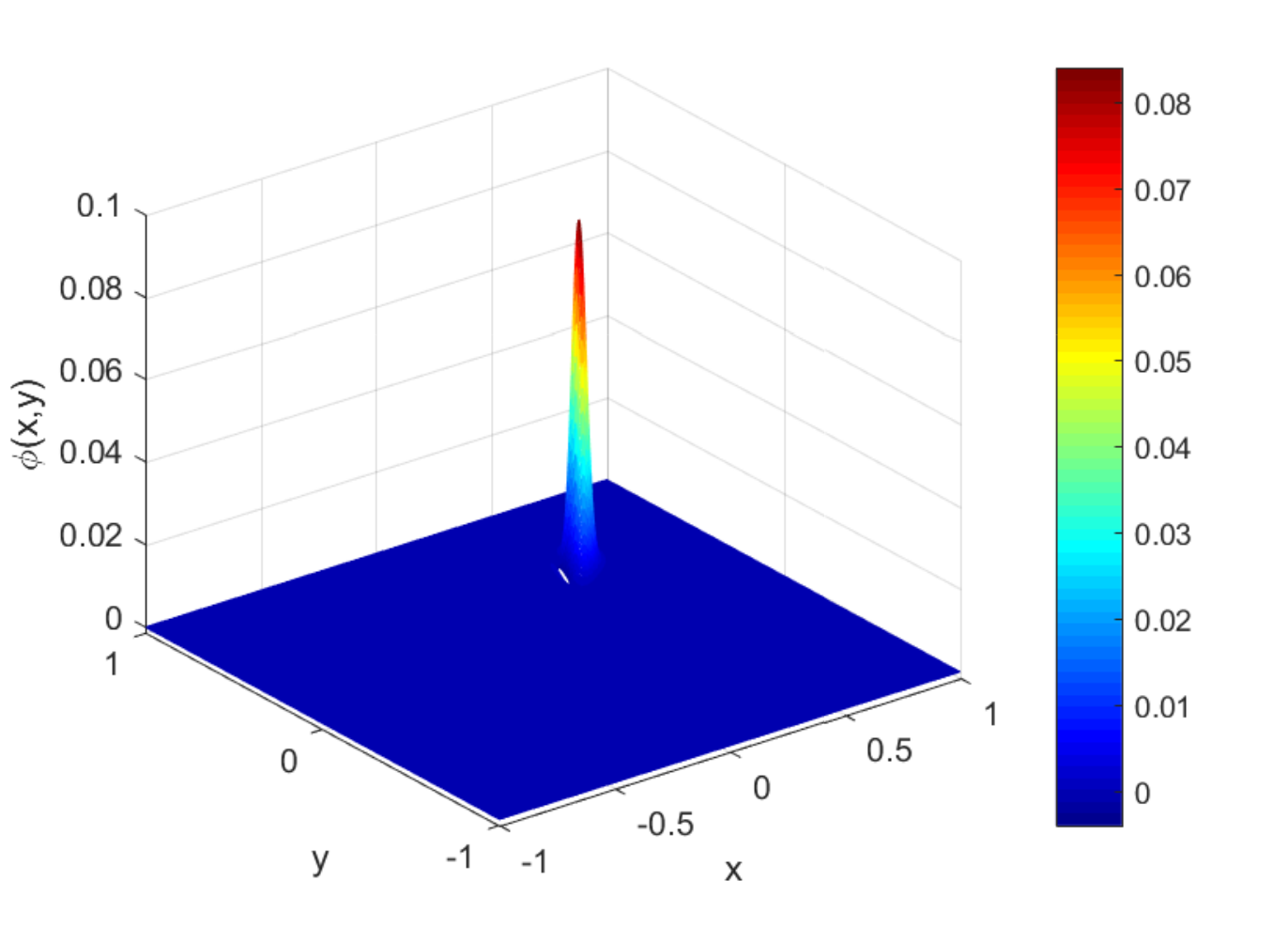}
      \label{Fig4_4_5_a}
    }
    \subfigure[]
    {
      \includegraphics[scale=0.5]{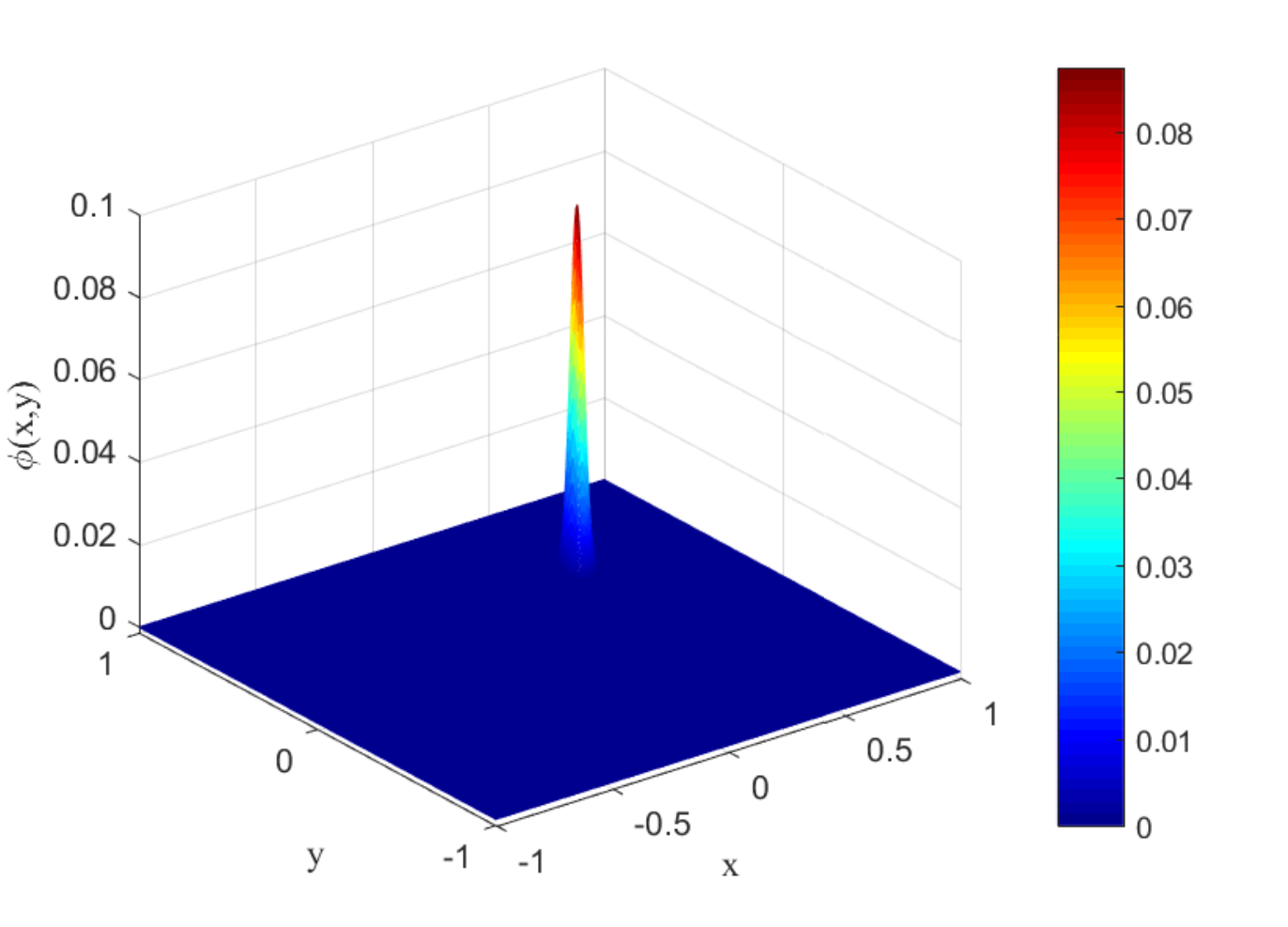}
      \label{Fig4_4_5_b}
    }
    \caption{Distributions of the scalar variable $\phi$ at time $t=5$ [fully anisotropic diffusion problem: (a) numerical solution, (b) analytical solution], where $\kappa=10^{-4}$, $\mathbf{u}=(0.1,0.1)$.}
    \label{Fig4_4_5}
  \end{figure}

  In the early work of Chai et al. \cite{chai2016multiple}, we have known that if $\kappa=10^{-4}$ and $\mathbf{u}=(0.1,0.1)^\intercal$, the SLBM could not give a stability solution. To test the stability of the present DUGKS, we performed some simulations with $\kappa=10^{-4}$ and $\mathbf{u}=(0.1,0.1)^\intercal$. As seen from Fig. \ref{Fig4_4_5}, the DUGKS can give a stable numerical and accurate solution.
  
  \textbf{Example 4.5} We continue to consider the following convection-diffusion equation with nonlinear convection and diffusion terms as
  \begin{equation}
    \partial_t \phi + \nabla \cdot (\phi^m \mathbf{u}) = \nabla \cdot [\alpha (\nabla \cdot \mathbf{D})] + F,
    \label{eq:4_5_1}
  \end{equation}
  where $m$ and $\alpha$ are two constants, $\mathbf{D}$ is the tensor function of $\phi$, and is given by
  \begin{equation}
    \mathbf{D} = \left(\begin{matrix}
      \phi^{n_x} \quad &0\\
      0 \quad &\phi^{n_y}
    \end{matrix}\right).
    \label{eq:4_5_2}
  \end{equation}
  $F$ is the source term, and is defined as
  \begin{equation}
    \begin{split}
      F&= \exp(-At) \{ A\cos(2\pi x)\cos(2\pi y)\\
      &-4n_x\pi^2\alpha\phi^{n_x-2}[(n_x-1)\exp(-At)\sin^2(2\pi x)\cos^2(2\pi y) + \phi \cos(2\pi x)\cos(2\pi y)]\\
      &-4n_y\pi^2\alpha\phi^{n_y-2}[(n_y-1)\exp(-At)\sin^2(2\pi y)\cos^2(2\pi x) + \phi \cos(2\pi x)\cos(2\pi y)]\\
      &+2\pi m\phi^{m-1}[u_x\sin(2\pi x)\cos(2\pi y) + u_y\cos(2\pi x)\sin(2\pi y)] \},
    \end{split}
    \label{eq:4_5_3}
  \end{equation}
  where $n_x$, $n_y$ and $A$ are the constants. Under the proper initial and periodic conditions, the analytical solution of this problem can be obtained,
  \begin{equation}
    \phi(x,y,t) = \kappa - \exp(-At)\cos(2\pi x)\cos(2\pi y).
    \label{eq:4_5_4}
  \end{equation}
  where $\kappa$ is a constant. For this problem, the functions $\mathbf{B}$ and $\mathbf{C}$ are given by
  \begin{equation}
    \mathbf{B} = (\phi^m u_x,\phi^m u_y)^\intercal, \quad \mathbf{C} = \frac{m^2\phi^{2m-1}}{2m-1}\left(\begin{matrix}
      u_x^2 &u_x u_y\\
      u_y u_x &u_y^2
    \end{matrix}\right).
    \label{eq:4_5_5}
  \end{equation}

\begin{figure}
    \subfigure[]
    {
      \includegraphics[scale=0.5]{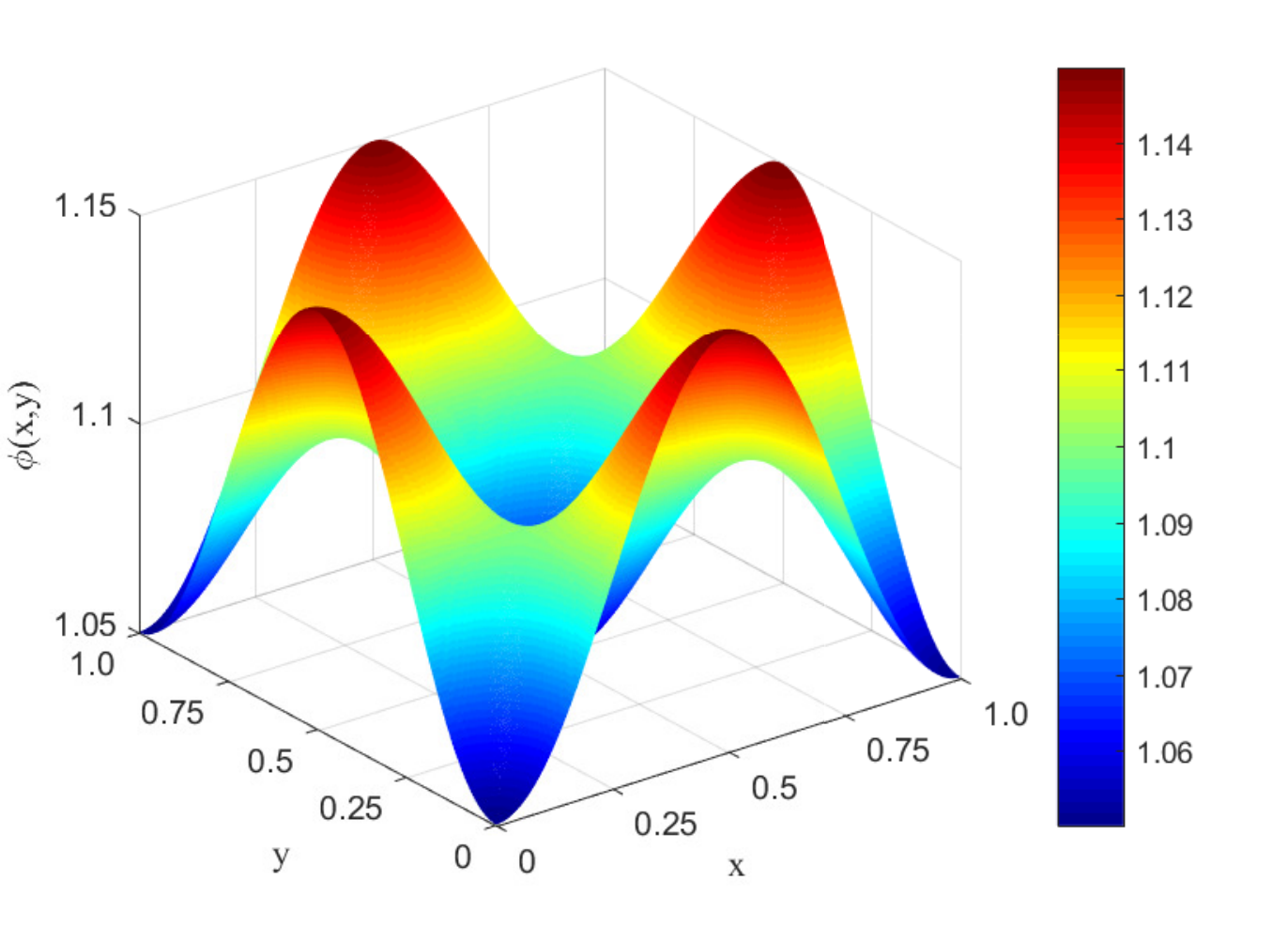}
      \label{Fig4_5_1_a}
    }
    \subfigure[]
    {
      \includegraphics[scale=0.5]{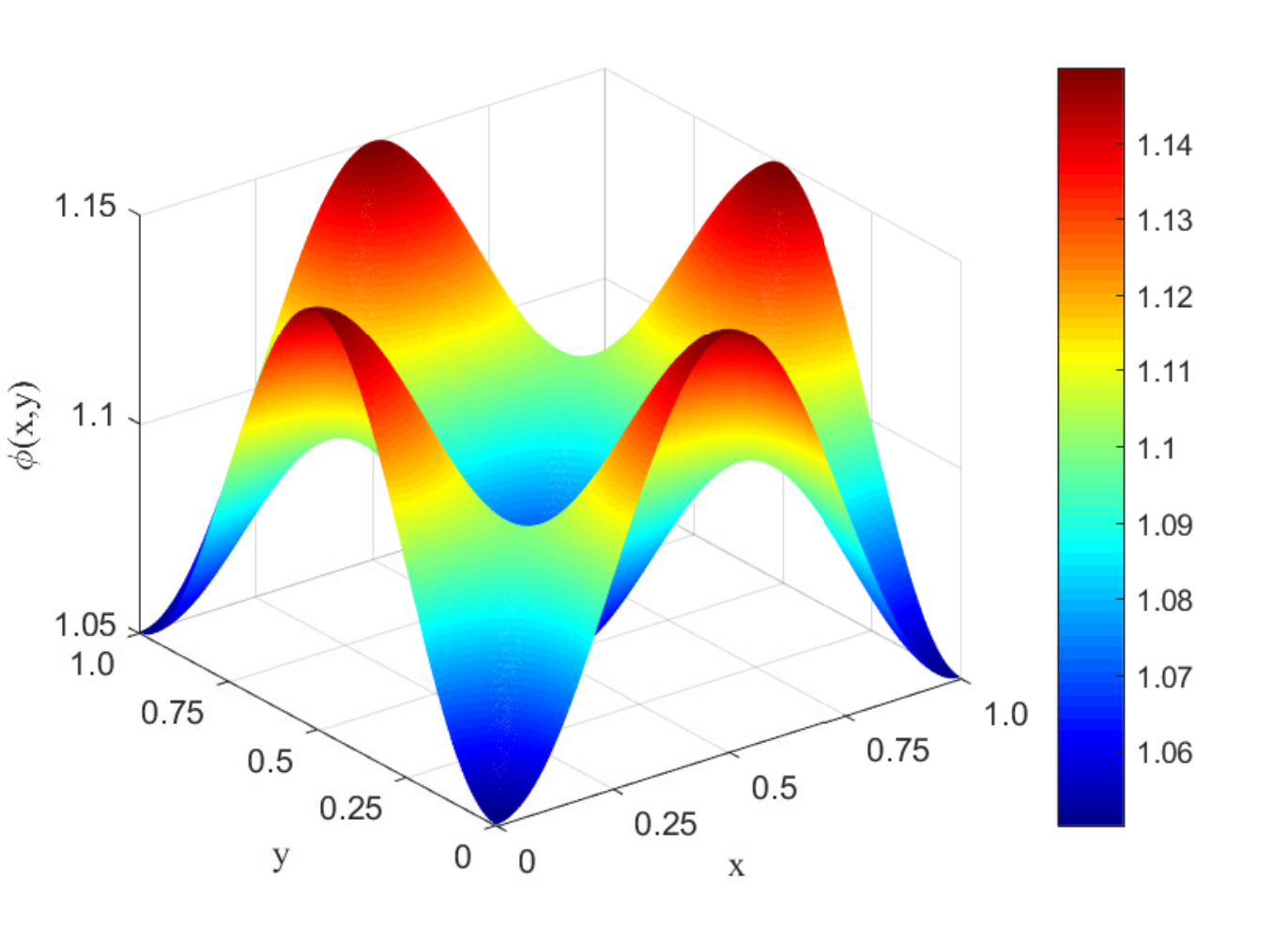}
      \label{Fig4_5_1_b}
    }
    \caption{Distributions of the scalar variable $\phi$ at $Pe = 100$ and $t = 3.0$ [(a) numerical solution, (b) analytical solution].}
    \label{Fig4_5_1}
  \end{figure}

\begin{figure}
    \subfigure[]
    {
      \includegraphics[scale=0.5]{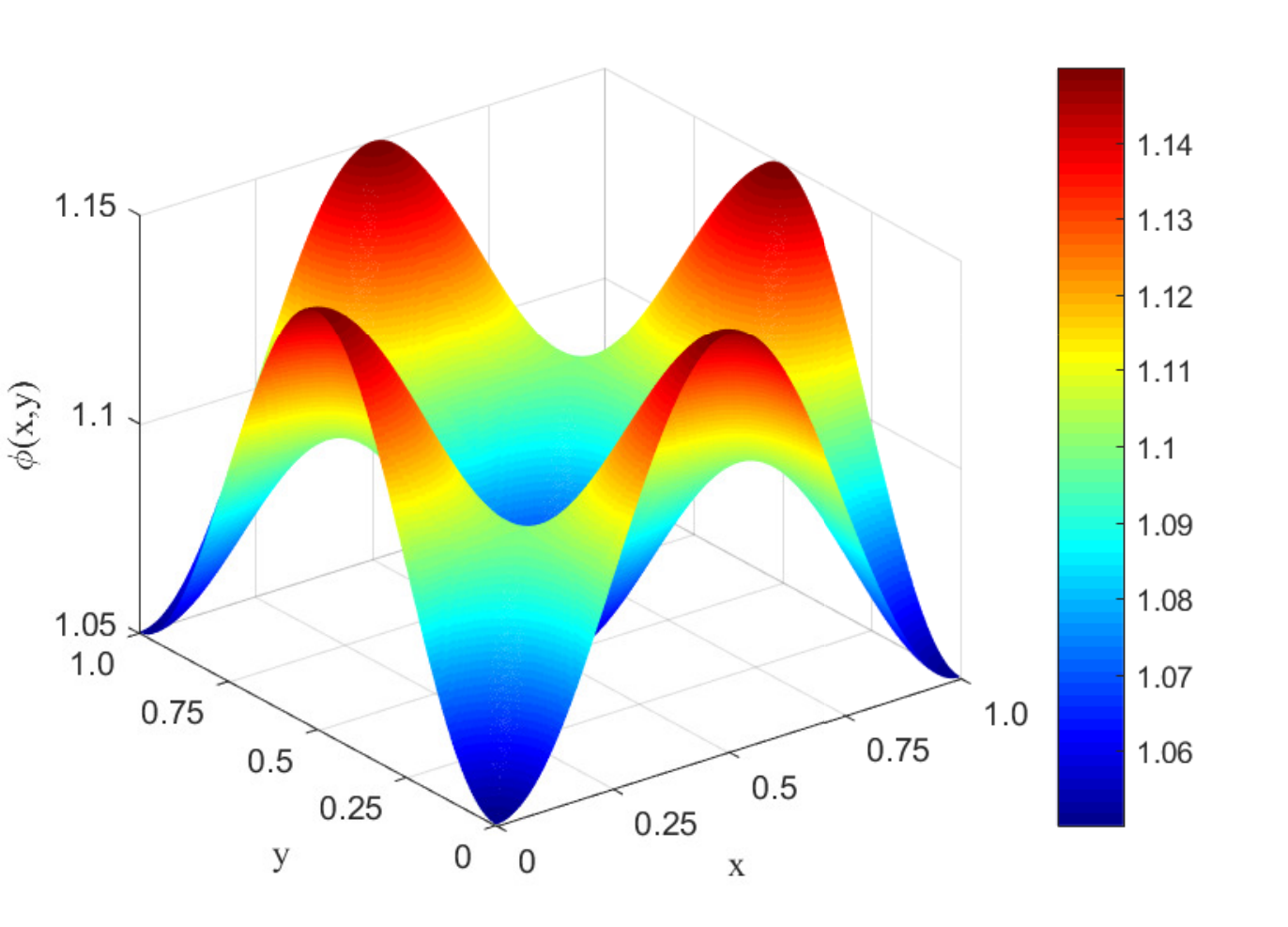}
      \label{Fig4_5_2_a}
    }
    \subfigure[]
    {
      \includegraphics[scale=0.5]{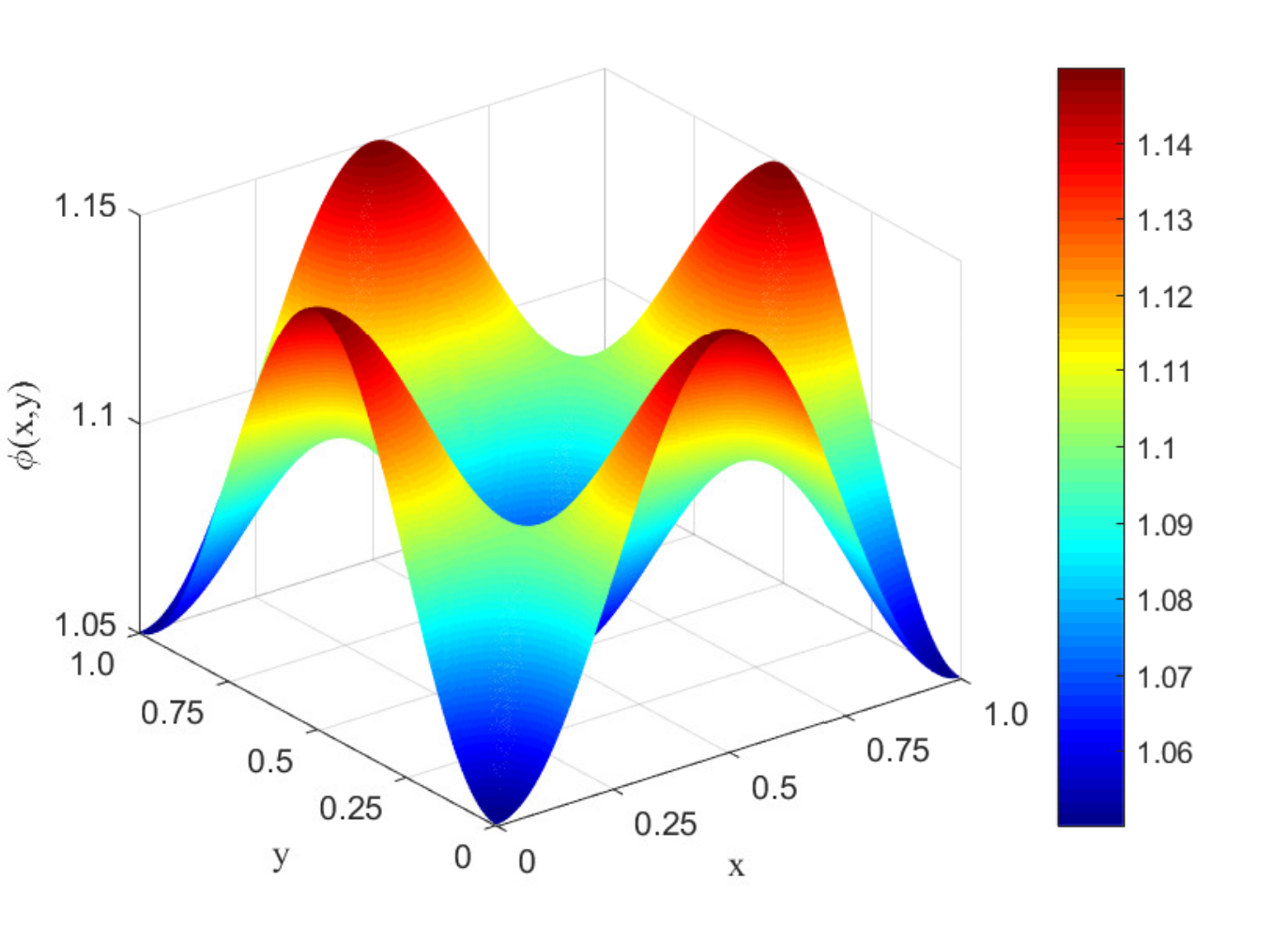}
      \label{Fig4_5_2_b}
    }
    \caption{Distributions of the scalar variable $\phi$ at $Pe = 1000$ and $t = 3.0$ [(a) numerical solution, (b) analytical solution].}
    \label{Fig4_5_2}
  \end{figure}

  We performed the simulations on [0,1] $\times$ [0,1] with the uniform grid 400 $\times$ 400, and the physical parameters are set as $\kappa = 1.1$, $A = 1.0$, $m = 2.0$, $n_x = 2.0$, $n_y = 3.0$ and $c = 1.0$. Besides, the CFL condition number is equal to 0.5. As seen from Figs. \ref{Fig4_5_1} and \ref{Fig4_5_2}, the numerical solutions at $t = 3.0$ and different P\'eclet numbers ($Pe = Lu_x/\alpha$) are in good agreement with analytical solutions, and the $GRE$s are about $3.348 \times 10^{-5}$ for $Pe = 100$ and $3.039 \times 10^{-5}$ for $Pe = 1000$. In addition, we also find that the values of $GRE$ are much smaller than $2.865 \times 10^{-3}$ and $7.162 \times 10^{-4}$ in Ref. \cite{chai2016multiple}.

\begin{figure}
    \centering
    \includegraphics[scale=0.7]{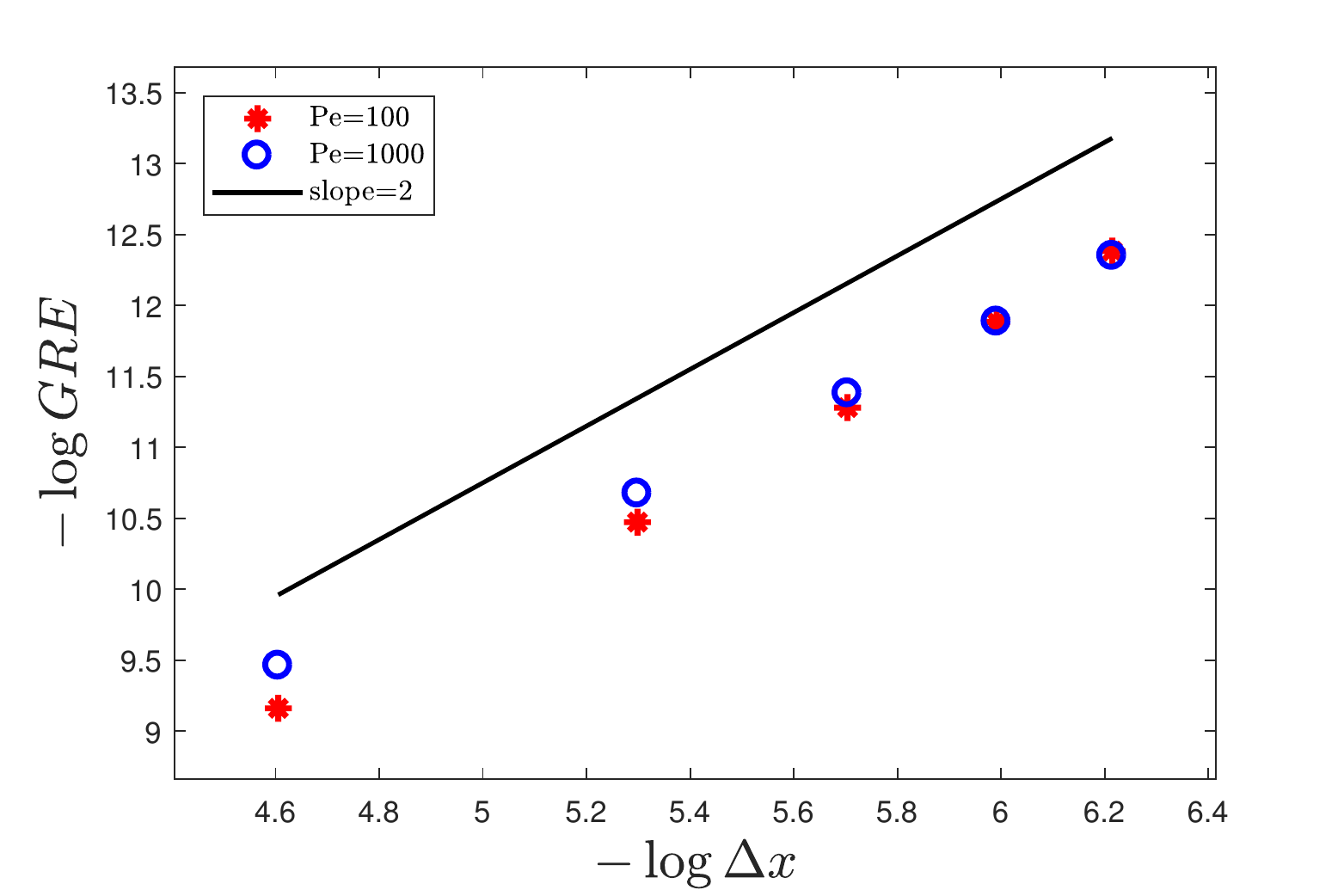}
    \caption{The global relative errors at different lattice sizes ($\Delta x=L/200,L/300,L/400,L/500,L/600$), the slope of the $solid line$ is $2.0$, which indicates the present DUGKS has a second-order convergence rate in space.}
    \label{Fig4_5_3}
  \end{figure}

  Then the convergence rate of the present DUGKS for this problem is also considered, and the lattice size is varied from 100 $\times$ 100 to 500 $\times$ 500 with a fixed time step fixed $\Delta t = 1.0\times 10^{-5}$. As shown in Fig. \ref{Fig4_5_3}, the present DUGKS also has a second-order convergence rate for this nonlinear convection-diffusion equation.

  \textbf{Example 4.6} We now considered the nonlinear Fokker-Planck equation \cite{wang2015regularized}
  \begin{equation}
    \frac{\partial \phi}{\partial t} + \frac{\partial \left\{[tx + \left\langle{x(t)}\right\rangle]\phi\right\}}{\partial x} = \frac{\partial^2(2t\phi)}{\partial x^2},
    \label{eq:4_6_1}
  \end{equation}
  with the initial condition
  \begin{equation}
    \phi(x,0) = \delta(x-1.0),
    \label{eq:4_6_2}
  \end{equation}
  and analytical solution
  \begin{equation}
    \phi(x,t) = \frac{1}{\sqrt{4\pi\eta(t)\exp(t^2)}} \exp\left\{-\frac{[x-\left\langle{x(t)}\right\rangle]^2}{4\eta(t)\exp(t^2)}\right\},
    \label{eq:4_6_3}
  \end{equation}
  where $\left\langle{x(t)}\right\rangle = \exp(t+\frac{t^2}{2})$, $\eta(t) = 1-\exp(-t^2)$, $\int_{-\infty}^{+\infty} \delta(x-x_0) = 1$ and
  \begin{equation*}
    \delta(x-x_0) = 
    \left\{
    \begin{array}{lr}
      \infty, \quad x = x_0,\\
      0, \quad x \neq x_0.
    \end{array}
    \right.
  \end{equation*}

  We noted that in the above tests, the convection term $\mathbf{B}$ is only a function of $\phi$. However, in this problem, the convection term $\mathbf{B}$ is the function of $\phi$, $x$ and $t$, thus we have to define the auxiliary moment $\mathbf{C}=0$. For this example, the initial condition of $\phi(x,0)$ is taken as
  \begin{equation}
    \phi(x_i,0)=
    \left\{
      \begin{array}{lr}
        \frac{1}{\Delta x}, \quad |x_i - x_0| \leq \eta,\\
        0, \quad \quad x \neq x_0,
      \end{array}
    \right.
    \label{eq:4_6_4}
  \end{equation}
  where $\eta$ is a small constant and $\Delta x$ is the lattice spacing.

\begin{figure}[ht]
    \centering
    \includegraphics[scale=0.7]{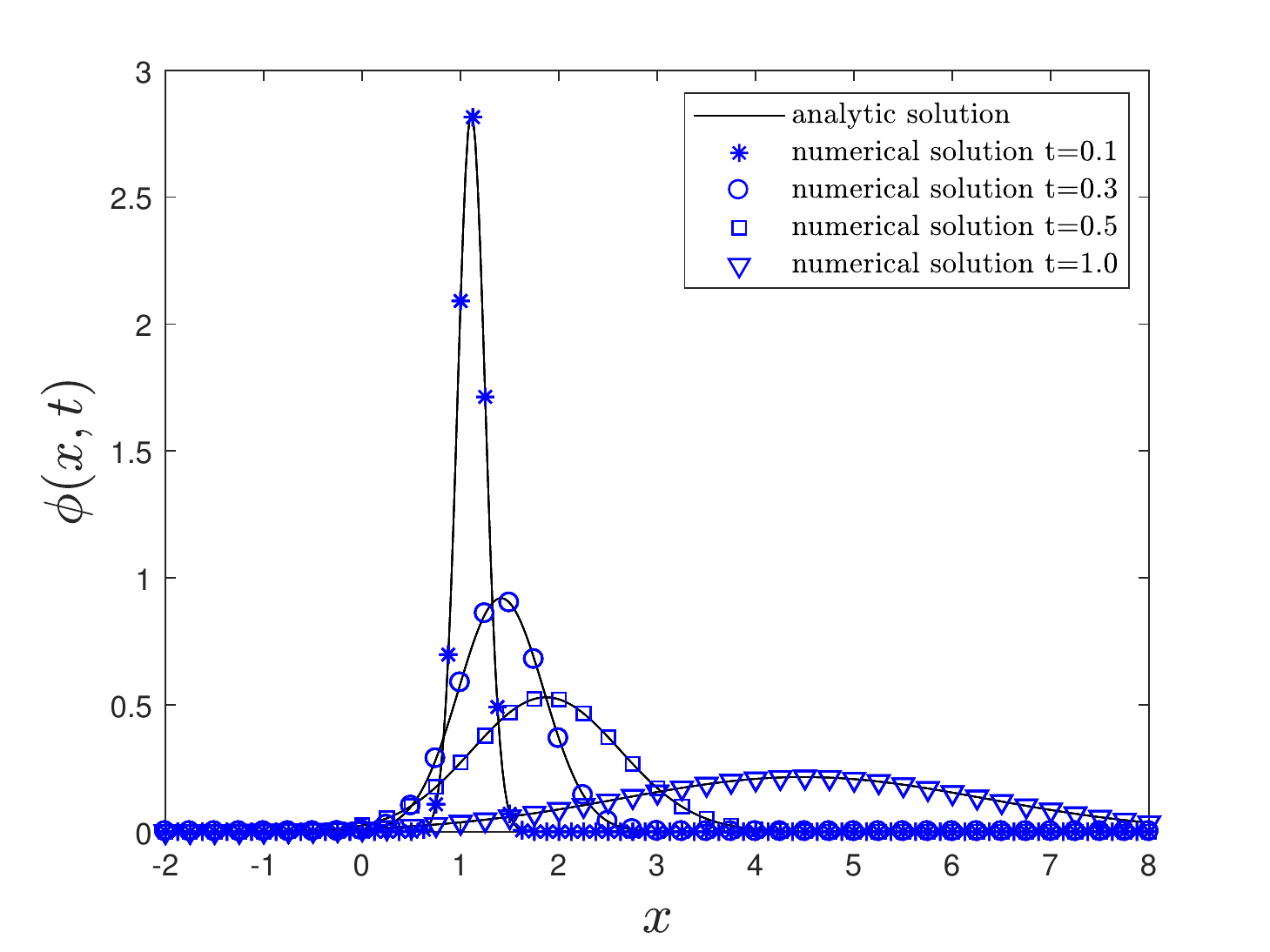}
    \caption{Analytical and numerical solutions of nonlinear Fokker-Planck equation at different time $t$}
    \label{Fig4_6_1}
  \end{figure}

  In our simulations, the physical domain is fixed on $[-2,8]$, the uniform grid 400 $\times$ 400 is adopted, $\text{CFL}=0.5$. We presented the results at different time in Fig. \ref{Fig4_6_1}. From this figure, we can see that the numerical solutions agree well with the analytical solutions. 

\begin{figure}
    \centering
    \includegraphics[scale=0.7]{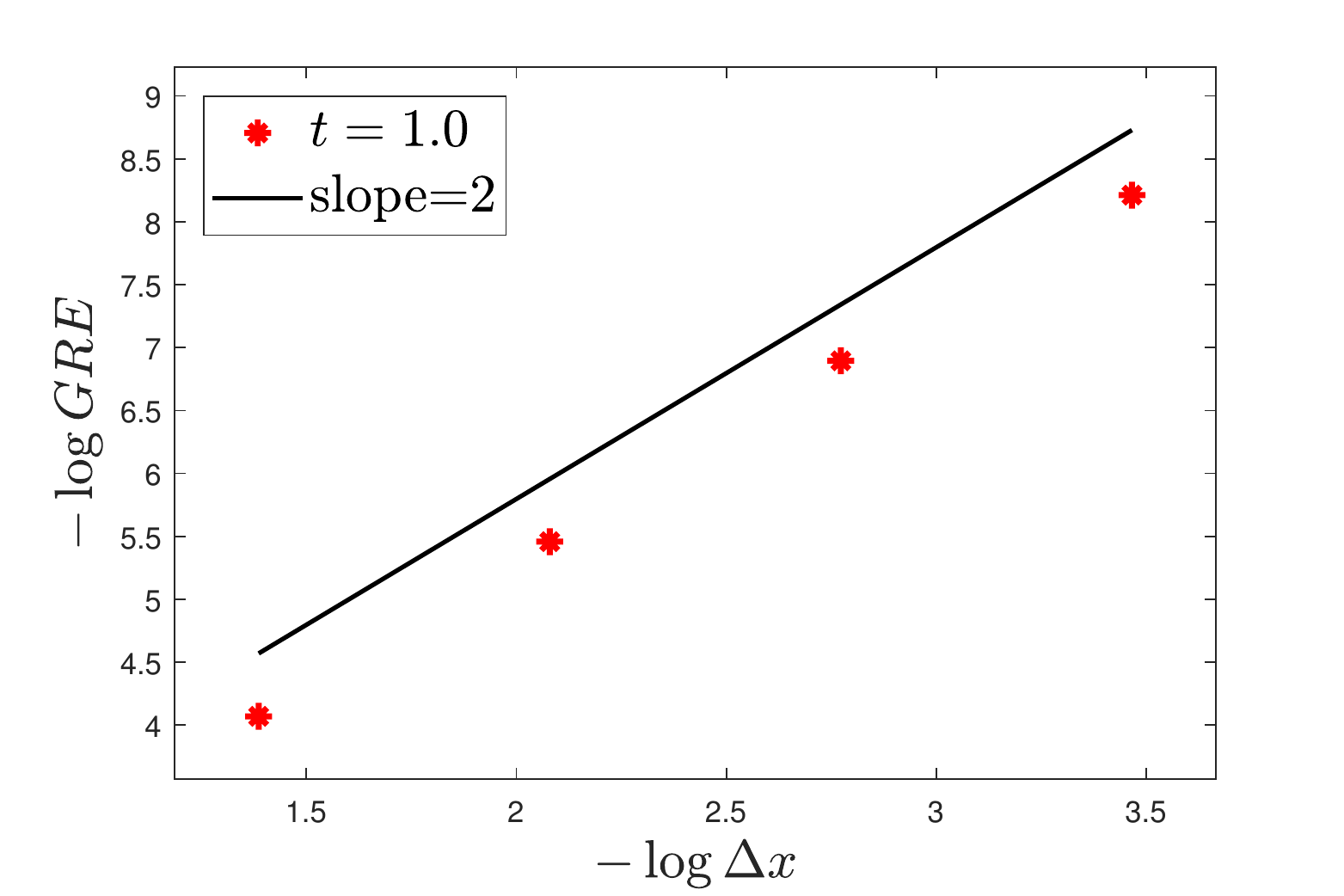}
    \caption{The global relative errors at $t = 1.0$ and different lattice sizes  ($\Delta x = 1/4 - 1/32$), the slope of the $solid line$ is 2.0, indicating the present DUGKS model has a second-order convergence rate in space.}
    \label{Fig4_6_2}
  \end{figure}

\begin{figure}
    \centering
    \includegraphics[scale=0.7]{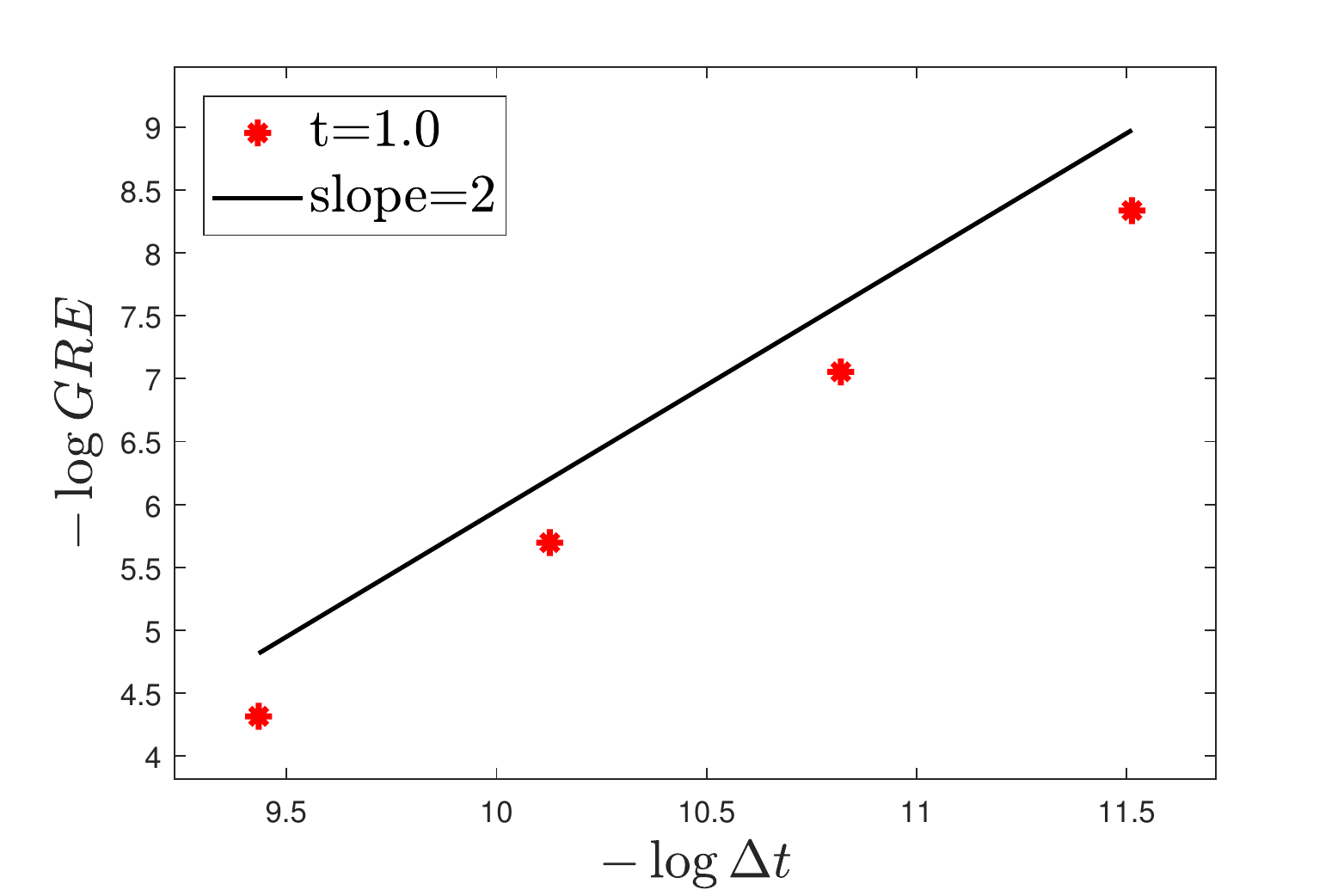}
    \caption{The global relative errors at $t = 1.0$ and different time step ($\Delta t = 8.0\times10^{-5} \sim 5.0\times10^{-6}$), the slope of the $solid line$ is 2.0, indicating the present DUGKS has a second-order convergence rate in time.}
    \label{Fig4_6_3}
  \end{figure}

  To test the convergence rate of the DUGKS for this problem, some simulations were carried out at different lattice size ($\Delta x = 1/4 \sim 1/32$), and the time step is fixed at $\Delta t = 1.0 \times 10^{-5}$. As shown in Fig. \ref{Fig4_6_2}, the present DUGKS indeed has a second-order convergence rate in space. Besides, theoretically, the DUGKS should also have a second-order convergence rate in time, to confirm this statement, we also carried out some simulations with a fixed $\Delta x = 1/100$, the time step is varied from $8.0\times10^{-5}$ to $1.0\times10^{-5}$. As seen from Fig. \ref{Fig4_6_3}, the present DUGKS does have a second-order convergence rate in time.

\section{\label{sec:level5}Conclusion}
  In this work, the discrete unified gas kinetic scheme is developed to solve general nonlinear convection-diffusion equation. Through Chapman-Enskog analysis, the NCDE can be recovered exactly from the present DUGKS. Through a lot of numerical simulations, we find that the numerical solutions are in good agreement with analytical solutions, and the present DUGKS has a second-order convergence rate in both space and time. In \textbf{Example 4.1}, a comparison was made between DUGKS, FV-LBM and MRT-LBM, the results show that the present DUGKS is more accurate than FV-LBM, and has almost same accuracy with the MRT-LBM. In \textbf{Example 4.2} and \textbf{Example 4.3}, one can see that the present DUGKS is efficient and can be implemented on the non-uniform meshes. In \textbf{Example 4.4}, the results show that the DUGKS model is more stable than SLBM. In \textbf{Example 4.5}, a more nonlinear equation is considered to test our model, and finally in \textbf{Example 4.6}, we tested the present DUGKS, and found that the DUGKS also has a second-order convergence rate in time.

\section*{Acknowledgements}
    This work is supported by the National Natural Science Foundation of China (Grants No. 51576079 and No.51836003), and the National Key Research and Development Program of China (Grant No. 2017YFE0100100)

\section*{References}

\bibliography{mybibfile}

\begin{thebibliography}{10}
\expandafter\ifx\csname url\endcsname\relax
  \def\url#1{\texttt{#1}}\fi
\expandafter\ifx\csname urlprefix\endcsname\relax\def\urlprefix{URL }\fi
\expandafter\ifx\csname href\endcsname\relax
  \def\href#1#2{#2} \def\path#1{#1}\fi

\bibitem{cussler2009diffusion}
E.~L. Cussler, Diffusion: mass transfer in fluid systems, Cambridge university
  press, 2009.

\bibitem{johnson2012numerical}
C.~Johnson, Numerical solution of partial differential equations by the finite
  element method, Courier Corporation, 2012.

\bibitem{thomas2013numerical}
J.~W. Thomas, Numerical partial differential equations: finite difference
  methods, Vol.~22, Springer Science \& Business Media, 2013.

\bibitem{leveque2002finite}
R.~J. LeVeque, Finite volume methods for hyperbolic problems, Vol.~31,
  Cambridge university press, 2002.

\bibitem{chen1998lattice}
S.~Chen, G.~D. Doolen, Lattice boltzmann method for fluid flows, Annu. Rev.
  Fluid Mech. 30~(1) (1998) 329--364.

\bibitem{guo2013lattice}
Z.~Guo, C.~Shu, Lattice Boltzmann method and its applications in engineering,
  Vol.~3, World Scientific, 2013.

\bibitem{succi2015lattice}
S.~Succi, Lattice boltzmann 2038, EPL 109~(5) (2015) 50001.

\bibitem{chen2014critical}
L.~Chen, Q.~Kang, Y.~Mu, Y.~He, W.~Tao, A critical review of the
  pseudopotential multiphase lattice boltzmann model: Methods and applications,
  Int. J. Heat Mass Transf. 76 (2014) 210--236.

\bibitem{dou2013numerical}
Z.~Dou, Z.~Zhou, Numerical study of non-uniqueness of the factors influencing
  relative permeability in heterogeneous porous media by lattice boltzmann
  method, Int. J. Heat Fluid Flow 42 (2013) 23--32.

\bibitem{chai2018comparative}
Z.~Chai, D.~Sun, H.~Wang, B.~Shi, A comparative study of local and nonlocal
  allen-cahn equations with mass conservation, Int. J. Heat Mass Transf. 122
  (2018) 631--642.

\bibitem{yuan2019dynamic}
X.~Yuan, Z.~Chai, B.~Shi, Dynamic behavior of droplet through a confining
  orifice: A lattice boltzmann study, Comput. Math. Appl. 77 (2019) 2640--2658.

\bibitem{ponce1993lattice}
S.~Ponce~Dawson, S.~Chen, G.~D. Doolen, Lattice boltzmann computations for
  reaction-diffusion equations, J. Chem. Phys. 98~(2) (1993) 1514--1523.

\bibitem{shi2009lattice}
B.~Shi, Z.~Guo, Lattice boltzmann model for nonlinear convection-diffusion
  equations, Phys. Rev. E 79~(1) (2009) 016701.

\bibitem{chopard2009lattice}
B.~Chopard, J.~L. Falcone, J.~Latt, The lattice boltzmann advection-diffusion
  model revisited, Eur. Phys. J.-Spec. Top. 171~(1) (2009) 245--249.

\bibitem{ginzburg2005equilibrium}
I.~Ginzburg, Equilibrium-type and link-type lattice boltzmann models for
  generic advection and anisotropic-dispersion equation, Adv. Water Resour.
  28~(11) (2005) 1171--1195.

\bibitem{ginzburg2005generic}
I.~Ginzburg, Generic boundary conditions for lattice boltzmann models and their
  application to advection and anisotropic dispersion equations, Adv. Water
  Resour. 28~(11) (2005) 1196--1216.

\bibitem{ginzburg2007lattice}
I.~Ginzburg, Lattice boltzmann modeling with discontinuous collision
  components: Hydrodynamic and advection-diffusion equations, J. Stat. Phys.
  126~(1) (2007) 157--206.

\bibitem{ginzburg2012truncation}
I.~Ginzburg, Truncation errors, exact and heuristic stability analysis of
  two-relaxation-times lattice boltzmann schemes for anisotropic
  advection-diffusion equation, Commun. Comput. Phys. 11~(5) (2012) 1439--1502.

\bibitem{ginzburg2013multiple}
I.~Ginzburg, Multiple anisotropic collisions for advection--diffusion lattice
  boltzmann schemes, Adv. Water Resour. 51 (2013) 381--404.

\bibitem{yoshida2010multiple}
H.~Yoshida, M.~Nagaoka, Multiple-relaxation-time lattice boltzmann model for
  the convection and anisotropic diffusion equation, J. Comput. Phys. 229~(20)
  (2010) 7774--7795.

\bibitem{chai2016multiple}
Z.~Chai, B.~Shi, Z.~Guo, A multiple-relaxation-time lattice boltzmann model for
  general nonlinear anisotropic convection--diffusion equations, J. Sci.
  Comput. 69~(1) (2016) 355--390.

\bibitem{guo2013discrete}
Z.~Guo, K.~Xu, R.~Wang, Discrete unified gas kinetic scheme for all knudsen
  number flows: Low-speed isothermal case, Phys. Rev. E 88~(3) (2013) 033305.

\bibitem{bhatnagar1954model}
P.~L. Bhatnagar, E.~P. Gross, M.~Krook, A model for collision processes in
  gases. i. small amplitude processes in charged and neutral one-component
  systems, Phys. Rev. 94~(3) (1954) 511.

\bibitem{wu2016discrete}
C.~Wu, B.~Shi, Z.~Chai, P.~Wang, Discrete unified gas kinetic scheme with a
  force term for incompressible fluid flows, Comput. Math. Appl. 71~(12) (2016)
  2608--2629.

\bibitem{zhangchunhua2018discrete}
C.~Zhang, K.~Yang, Z.~Guo, A discrete unified gas-kinetic scheme for immiscible
  two-phase flows, Int. J. Heat Mass Transf. 126 (2018) 1326--1336.

\bibitem{yang2019phase}
Z.~Yang, C.~Zhong, C.~Zhuo, Phase-field method based on discrete unified
  gas-kinetic scheme for large-density-ratio two-phase flows, Phys. Rev. E
  99~(4) (2019) 043302.

\bibitem{cahn1958free}
J.~W. Cahn, J.~E. Hilliard, Free energy of a nonuniform system. i. interfacial
  free energy, J. Chem. Phys. 28~(2) (1958) 258--267.

\bibitem{cahn1959free}
J.~W. Cahn, J.~E. Hilliard, Free energy of a nonuniform system. iii. nucleation
  in a two-component incompressible fluid, J. Chem. Phys. 31~(3) (1959)
  688--699.

\bibitem{geier2015conservative}
M.~Geier, A.~Fakhari, T.~Lee, Conservative phase-field lattice boltzmann model
  for interface tracking equation, Phys. Rev. E 91~(6) (2015) 063309.

\bibitem{huo2018discrete}
Y.~Huo, Z.~Rao, The discrete unified gas kinetic scheme for solid-liquid phase
  change problem, International Communications in Heat and Mass Transfer 91
  (2018) 187--195.

\bibitem{patil2013chapman}
D.~Patil, Chapman--enskog analysis for finite-volume formulation of lattice
  boltzmann equation, Physica A 392~(12) (2013) 2701--2712.

\bibitem{ubertini2004recent}
S.~Ubertini, S.~Succi, Recent advances of lattice boltzmann techniques on
  unstructured grids, Prog. Comput. Fluid Dyn. 5~(1-2) (2004) 85--96.

\bibitem{wang2015regularized}
L.~Wang, B.~Shi, Z.~Chai, Regularized lattice boltzmann model for a class of
  convection-diffusion equations, Phys. Rev. E 92~(4) (2015) 043311.

\end{thebibliography}

\end{document}